\documentclass[12pt, a4paper]{article}
\usepackage[font=small,format=plain,labelfont=bf,up,textfont=normal,up,justification=justified,singlelinecheck=false]{caption}
\usepackage{subcaption}
\usepackage{mathrsfs}
\usepackage[a4paper, left=2cm,right=2cm]{geometry}
\usepackage[colorlinks=true,linkcolor=black,citecolor=teal,urlcolor=MidnightBlue,filecolor=black]{hyperref}
\usepackage{amsfonts}
\usepackage{amsmath,amssymb}
\usepackage{setspace}
\usepackage{slashed}
\usepackage{braket}
\usepackage[dvipsnames]{xcolor}
\definecolor{SchoolColor}{rgb}{0.6471, 0.1098, 0.1882}
\usepackage{subfloat}
\usepackage[utf8,applemac]{inputenc}
\usepackage{tensor}
\usepackage{cite}
\usepackage{tikz}
\usepackage[thicklines]{cancel}
\usetikzlibrary{calc}
\usetikzlibrary{patterns}
\usetikzlibrary{arrows.meta}
\usepackage{graphicx}
\usepackage{bm}
\allowdisplaybreaks[4]
\usepackage[utf8,applemac]{inputenc}
\usepackage{tensor}
\usepackage{cite}
\usepackage{tikz}
\usepackage{graphicx}
\graphicspath{{figure/}}
\usepackage{array}
\usepackage{booktabs}
\usepackage{multirow}
\usepackage{dcolumn}
\usepackage{bm}
\usepackage{verbatim}
\usepackage{textcomp}
\usepackage{graphicx} 

\numberwithin{equation}{section}
\newcommand{\bea}{\begin{eqnarray}}
\newcommand{\eea}{\end{eqnarray}}
\newcommand{\be}{\begin{equation}}
\newcommand{\ee}{\end{equation}}
\newcommand{\bs}{\begin{subequations}}
\newcommand{\es}{\end{subequations}}
\def\nn{\nonumber}

\def\p{\partial}

\newcommand{\n}{\nabla }
\newcommand{\beqs}{\begin{eqnarray}}
\newcommand{\eeqs}{\end{eqnarray}}
\numberwithin{equation}{section}

\newcommand{\Rmnum}[1]{\uppercase\expandafter{\romannumeral #1\relax}}

\def\ch{\mathcal{H}}\def\ci{\mathcal{I}}\def\cl{\mathcal{L}}\def\cm{\mathcal{M}}\def\co{\mathcal{O}}\def\ct{\mathcal{T}}

\def\c.c.{\mathrm{c.c.}}
\def\mn{{\mu\nu}}
\def\a{\alpha}

\def\ep{\epsilon}

\def\m{\mu}\def\n{\nu}

\def\vp{\varphi}
\def\Om{\Omega}

\begin{document}
\begin{titlepage}

\begin{flushright}\vspace{-3cm}
{\small\today }
\end{flushright}
\vspace{0.5cm}
\begin{center}
	{{ \LARGE{\bf{Electromagnetic helicity flux operators \vspace{8pt}\\ in higher dimensions }}}}\vspace{5mm}
	
	\centerline{\large{Wen-Bin Liu\footnote{liuwenbin0036@hust.edu.cn}, Jiang Long\footnote{longjiang@hust.edu.cn} \& Xin-Hao Zhou\footnote{zhouxinhao@hust.edu.cn}}}
	\vspace{2mm}
	\normalsize
	\bigskip\medskip
	\textit{School of Physics, Huazhong University of Science and Technology, \\ Luoyu Road 1037, Wuhan, Hubei 430074, China
	}
	\vspace{25mm}
	
	\begin{abstract}
		\noindent
		{The helicity flux operator is a fascinating quantity that characterizes the angular distribution of the helicity of radiative photons or gravitons and it has many interesting physical consequences. In this paper, we construct the electromagnetic helicity flux operators which form a non-Abelian group in general dimensions, among which the minimal helicity flux operators form the massless representation of the little group, a finite spin unitary irreducible representation of the Poincar\'e group. As in four dimensions, they generate an extended  angle-dependent transformation on the Carrollian manifold. 
		Interestingly, there is no known corresponding bulk duality transformation in general dimensions. However, we can construct a topological Chern-Simons term that evaluates the minimal helicity flux operators at $\mathcal{I}^+$.}
	\end{abstract}
\end{center}

\end{titlepage}
\tableofcontents

\section{Introduction}
Waves carry energy, momentum, and angular momentum during their propagation. The configuration of the wave is complicated in the near zone where the source is located. However, one could expect the situation to get simplified as it radiates to future null infinity $\mathcal{I}^+$, which is known as a Carrollian manifold in the literature \cite{Une,Gupta1966OnAA,Duval_2014a,Duval_2014b,Duval:2014uoa}. The radiation data is encoded in $\mathcal{I}^+$, and characterized by a fundamental field,  
the leading fall-off coefficient of the asymptotic expansion of the bulk field in four-dimensional asymptotically flat spacetime. The radiation fluxes from bulk to boundary could be constructed by the fundamental field, extending the famous mass loss formula \cite{Bondi:1962px}.

The fundamental field should be quantized due to the nature of the microscopic world. This has been studied in  \cite{Liu:2022mne,Liu:2024nkc} using the method of bulk reduction, where the commutators of the fundamental fields are consistent with the ones using asymptotic symplectic quantization \cite{Frolov:1977bp,1978JMP....19.1542A,Ashtekar:1981sf,Ashtekar:1981bq,Ashtekar:1987tt},  while need modifications at the boundary of $\ci^\pm$ \cite{Strominger:2013lka,Strominger:2013jfa,He:2014laa,He:2014cra,Strominger:2017zoo}. Interestingly, the extended Poincar\'e fluxes, which are interpreted as supertranslation and superrotation generators, form a closed algebra \cite{Liu:2022mne} which is the infinitesimal version of Carrollian diffeomorphism \cite{Ciambelli:2018xat}. The Carrollian diffeomorphism is deformed quantum-mechanically due to the appearance of a central charge in the algebra. This systematic treatment has been extended to the  vector theory \cite{Liu:2023qtr}, gravitational theory \cite{Liu:2023gwa} and higher spin theories \cite{Liu:2023jnc}. One of the important discoveries is the unexpected helicity flux operator in the theory with a non-vanishing spin. The helicity flux operator deforms the Carrollian diffeomorphism to the so-called intertwined Carrollian diffeomorphism and characterizes the angular distribution of the difference  between the numbers of massless particles with left and right hand helicities. It corresponds to the duality invariance of the theory \cite{Dirac:1931kp,Deser:1976iy,Henneaux:2004jw} in the bulk, which generates the superduality transformation of the fundamental field at the null boundary. There are various discussions on the duality transformation and its physical consequences \cite{Godazgar:2018qpq,Godazgar:2019dkh,Godazgar:2020kqd,Oliveri:2020xls,Freidel:2021fxf,Freidel:2021qpz,Seraj:2022qyt,Oblak:2023axy}. The helicity flux is an interesting physical observable, and it is equally important as the Poincar\'e fluxes. Recently, a quadrupole formula for  gravitational helicity flux density has been derived, and it has been used to investigate two-body systems \cite{Dong:2024ily}. It is expected to be studied systematically in the framework of post-Newtonian expansion \cite{2014LRR....17....2B}.

Interestingly, in the literature on the radiative phase space and asymptotic symmetry in higher dimensions \cite {Kapec:2014zla,Kapec:2015vwa,Mao:2017wvx,Campiglia:2017xkp,Pate:2017fgt,Campoleoni:2017qot,He:2019jjk,He:2019pll,Campoleoni:2020ejn,Capone:2021ouo}, the helicity flux operator has not been reported so far.
Recall that the method of bulk reduction has been generalized to higher dimensions,  and the Carrollian diffeomorphism is realized for the scalar theory \cite{Li:2023xrr}, bypassing various difficulties in the traditional asymptotic symmetry analysis \cite{Hollands:2003ie,Hollands:2004ac,Tanabe:2011es}. A generalization of theories with non-zero spin calls for an analogous helicity flux operator in higher dimensions. However, there is no known duality invariance of the bulk massless theory in higher dimensions. As a consequence, there is an obvious difficulty to construct the corresponding ``duality'' current in the bulk. This paper aims to deal with this tension. We propose that the transformation law of the fundamental field on the Carrollian manifold is completely the same as the one in four dimensions, and this point could be checked by bulk reduction for the vector theory. The supertranslation and superrotation generators are constructed respectively. They still form a closed algebra which is a higher dimensional extension of the intertwined Carrollian  diffeomorphism after including a helicity flux operator. The new operator is shown to count the number difference for the massless particles with left and right hand helicities associated with a fixed rotation plane in the mode expansion.  Since there are more than two transverse directions, the helicity flux operators generate a non-Abelian group, extending the Abelian group in four dimensions.

The layout of the paper is as follows. We will construct the electromagnetic helicity flux operator in higher dimensions from the closure of the  algebra in section \ref{helicityflux}. This is found by carefully studying the Hamilton's equations and the commutators of the Hamiltonians at the boundary. The results have been checked by bulk reduction in section \ref{bulk}.  We interpret the helicity flux operator using mode expansion in section \ref{int} and discuss its relation to topological Chern-Simons term in Carrollian manifold in the following section.  In section \ref{alt}, we discuss the helicity flux operator from different aspects.  We will conclude in section \ref{conc}. Various identities, computation of commutators, vector field  on a general null hypersurface and the helicity representation of the Poincar\'e group have been separated into several appendices.

\section{Boundary aspects}\label{helicityflux}
In this section, we will use the intrinsic method to derive the helicity flux operator where the input information is just the assumption that the  boundary symplectic form and the covariant variation of the boundary field under Carrollian diffeomorphism are the same as the ones in four dimensions. 

\subsection{Carrollian diffeomorphism}
 In this work, we will focus on $d$-dimensional  Minkowski spacetime which can be described by Cartesian coordinates $x^\mu=(t,x^i)$, where $\mu=0,1,2,\cdots,d-1$ are spacetime coordinates and $i=1,2,\cdots,d-1$ are spatial coordinates. We can also  use the spherical coordinates $(r,\Omega)$ to cover the  $(d-1)$-dimensional Euclidean space in which the radial distance $r$ is defined as usual \bea 
 r=\sqrt{x^i x^i},
 \eea and the angular  coordinates are collected as 
 \bea 
 \Omega=(\theta^1,\theta^2,\cdots,\theta^{m}),\quad m=d-2.
 \eea We will use the capital Latin alphabet $A,B,\cdots=1,2,\cdots,m$ to denote the components of the angular coordinates. In retarded coordinates $x^\alpha=(u,r,\theta^A)$, the metric of the Minkowski spacetime reads
 \bea 
 ds^2=-du^2-2du dr+r^2\gamma_{AB}d\theta^A d\theta^B,
 \eea where the retarded time is $u=t-r$ and the metric of the unit sphere $S^{m}$ reads 
\begin{align}\label{metricgamma}
  \gamma_{AB}=
  \begin{pmatrix}
    1&0&0&\cdots&0\\
    0&\sin^2\theta_1&0&\cdots&0\\
    0&0&\sin^2\theta_1\sin^2\theta_2&\cdots&0
    \\
    \vdots&\vdots&\vdots&\ddots&\vdots\\
    0&0&0&\cdots&\sin^2\theta_1\cdots\sin^2\theta_{m-1}
  \end{pmatrix}.
\end{align} 
The future null infinity $\mathcal{I}^+$ is a $(d-1)$-dimensional Carrollian manifold with a degenerate metric
 \bea 
 ds^2_{\mathcal{I}^+}=\gamma_{AB}d\theta^A d\theta^B,\label{degemet}
 \eea 
and a null vector 
\be 
\bm\chi=\partial_u
\ee which is to generate the retarded time direction. Moreover, the null vector is the kernel of the metric.

Carrollian diffeomorphism is generated by the vector $\bm\xi$ which preserves the null structure of the Carrollian manifold \cite{Ciambelli:2018xat,Ciambelli:2018wre,Ciambelli:2019lap} 
\bea 
\mathcal{L}_{\bm\xi}\bm\chi\propto\bm\chi
\eea whose solution is 
\bea 
\bm\xi=f(u,\Omega)\partial_u+Y^A(\Omega)\partial_A.
\eea The vector $\bm\xi$ may be separated into two parts 
\bea 
\bm\xi=\bm\xi_f+\bm\xi_{\bm Y}
\eea where $\bm\xi_f$ is parameterized by a smooth function $f(u,\Omega)$ on the Carrollian manifold
\bea 
\bm\xi_f=f(u,\Omega)\partial_u,\label{stv}
\eea and $\bm\xi_{\bm Y}$ is parameterized by a smooth vector $Y^A(\Omega)$ on the unit sphere $S^{m}$ 
\bea 
\bm\xi_{\bm Y}=Y^A(\Omega)\partial_A.\label{srv}
\eea The transformations generated by $\bm\xi_f$ and $\bm\xi_{\bm Y}$ are called general supertranslation and special superrotation, respectively \footnote{In the terminology of \cite{Liu:2022mne}, a time-dependent $f(u,\Omega)$ corresponds to a general supertranslation and a time-dependent $Y^A(u,\Omega)$ corresponds to a general superrotation. The general supertranslation $f(u,\Omega)$ can be expanded as a series of $u$ 
\be 
f(u,\Omega)=T(\Omega)+u B(\Omega)+\cdots\nn
\ee where a time-independent $T(\Omega)$ corresponds to the standard supertranslation \cite{Bondi:1962px} and $B(\Omega)$ is related to the boost. A time-independent $Y^A(\Omega)$ that is smooth on the sphere is called a special superrotation which is actually the superrotation in the generalized BMS algebra \cite{Campiglia:2014yka, Campiglia:2015yka}. On the other hand, a time-independent $Y^A(\Omega)$ that is a local conformal Killing vector corresponds to the superrotation in the extended BMS algebra \cite{Barnich:2010eb}. }. In this work, we just call them supertranslation and superrotation for brevity.

\subsection{Hamilton's equation}
We need to define a natural boundary field on Carrollian manifold $\mathcal{I}^+$. In our previous discussions \cite{Liu:2024nkc}, this is found by embedding the Carrollian manifold into the higher dimensional Lorentz manifold where the quantum field theory is well defined and then reducing the bulk theory to the boundary Carrollian manifold. The embedding of the Carrollian manifold in an ambient space can be found in \cite{deBoer:2003vf}. See also the recent discussion on this point in \cite{Salzer:2023jqv}.  

The lesson from our previous bulk reduction formalism is that the boundary metric should be kept invariant. Another intriguing property is that the leading boundary field is non-dynamical to  encode the full radiative data of the bulk theory. Nevertheless, there is a symplectic form to quantize the boundary field. The symplectic form shares the same form for general dimensions \cite{Li:2023xrr}.  Therefore, it is natural to impose the following symplectic form for higher dimensional electromagnetic theory 
\bea 
\bm\Omega(\delta A;\delta A;A)=\int du d\Omega\, \delta A_A\wedge \delta \dot{A}^A, \label{symp1}
\eea where $A_A$ is the fundamental field. From the symplectic form,  we may obtain the fundamental commutation relation 
\bea 
[A_A(u,\Omega),A_B(u',\Omega')]=\frac{i}{2}\gamma_{AB}\alpha(u-u')\delta(\Omega-\Omega'),\label{comAA}
\eea where the Dirac function on the sphere reads out explicitly as
\be 
\delta(\Omega-\Omega')=\frac{1}{\sqrt{\gamma}}\delta(\theta_1-\theta_1')\cdots \delta(\theta_{m}-\theta'_{m}),\label{diracsphere}
\ee 
and the function $\alpha(u-u')$ is defined as
\bea 
\alpha(u-u')=\frac{1}{2}[\theta(u'-u)-\theta(u-u')].
\eea 
For a general variation of the boundary field $A_A$ generated by $\bm\xi$, Hamilton's equation is written as 
\bea 
\delta H_{\bm\xi}=i_{\bm\xi}\bm\Omega.\label{hamiltonian}
\eea 

It has been shown that the  transformation law of the boundary field under Carrollian diffeomorphism 
is independent of the dimension for the scalar field \cite{Li:2023xrr}
\bea 
\slashed\delta_f\Sigma&=&\Delta(f;\Sigma;u,\Omega)=f(u,\Omega)\dot{\Sigma}(u,\Omega),\\
\slashed\delta_{\bm Y}\Sigma&=&\Delta(\bm Y;\Sigma;u,\Omega)=Y^A\nabla_A\Sigma+\frac{1}{2}\nabla_CY^C \Sigma.
\eea This is due to the intrinsic property of the Carrollian manifold. The symbol $\slashed\delta$ denotes the so-called covariant variation which was firstly defined in \cite{Liu:2023qtr}. Note that this delta slash is different from the one introduced in the formalism of covariant phase space \cite{Barnich:2001jy,Barnich:2011mi} where the delta slash is to mention that the infinitesimal variation of the charge is not necessarily integrable.  Therefore, we may assume the following variations of the vector field $A_A$ under supertranslation and superrotation 
\bea 
\slashed\delta_f A_A&\equiv&\Delta_A(f;A;u,\Omega)=f(u,\Omega)\dot{A}_A(u,\Omega),\label{st}\\
\hspace{-2em}  \slashed\delta_{\bm Y}A_A&\equiv&\Delta_A(Y;A;u,\Omega)=Y^C\nabla_CA_A+\frac{1}{2}\nabla_CY^C A_A+\frac{1}{2}(\nabla_AY_C-\nabla_CY_A)A^C.\label{sr}
\eea The assumption is a bit ad hoc. However, we will derive it from bulk reduction later.
In the first line, the transformation of the field $A_A$ under supertranslation may be derived from the Lie derivative of the $A_A$ along the direction of $\bm\xi_f$
\bea 
\mathcal{L}_{\bm\xi_f}A_A=f(u,\Omega)\dot{A}_A(u,\Omega).
\eea Note that for supertranslation, the Lie derivative of the scalar field $\Sigma$ along $\bm\xi_f$ is also  the same as the covariant variation
\bea 
\mathcal{L}_{\bm\xi_f}\Sigma=f(u,\Omega)\dot{\Sigma}(u,\Omega)=\Delta(f;\Sigma;u,\Omega).
\eea 
In the second line, the transformation of the field $A_A$ under superrotation does not coincide with the Lie derivative of $A_A$ along the direction of $\bm\xi_{\bm Y}$
\bea 
\mathcal{L}_{\bm\xi_{\bm Y}}A_A=Y^C\nabla_CA_A+\nabla_AY^C A_C\not=\Delta_A(Y;A;u,\Omega).\label{covYA}
\eea

Combining Hamilton's equation and the covariant variation, we find the following two quantities
\bs\begin{align} 
H_f&=\int du d\Omega\dot{A}^A\Delta_A(f;A;u,\Omega),\\
H_{\bm Y}&=\int du d\Omega\dot{A}^A\Delta_A(Y;A;u,\Omega).
\end{align}\es The results agree with the conclusion that \cite{Liu:2023gwa}
\bea 
H_{\bm\xi}=\int du d\Omega \dot{F}\slashed\delta_{\bm\xi}F
\eea is the general Hamiltonian for Carrollian diffeomorphism $\bm\xi$ in terms of the fundamental field $F$. After some effort, we may rewrite the two quantities in the following form
\bs\begin{align}
\mathcal{T}_f&=\int du d\Omega f(u,\Omega):\dot{A}^A\dot{A}_A:,\\
\mathcal{M}_{\bm Y}&=\frac{1}{2}\int du d\Omega Y^A(:\dot{A}^B\nabla^CA^D-A^B\nabla^C\dot{A}^D:)P_{ABCD}\label{MY}
\end{align}\es with 
\bea 
P_{ABCD}=\gamma_{AB}\gamma_{CD}+\gamma_{AC}\gamma_{BD}-\gamma_{AD}\gamma_{BC}.
\eea  We have added the normal-ordering symbol $:\cdots:$ in above expressions, and used a boldface letter $\bm Y$ in the operator $\mathcal{M}_{\bm Y}$ to emphasize that $\bm Y$ is a vector on the unit sphere. 
As a consequence, we find the following commutation relations 
\bs\begin{align}
\ [\mathcal{T}_f,A_A(u,\Omega)]&=-i\Delta_A(f;A;u,\Omega)=-i\slashed\delta_f A_A,\\
\ [\mathcal{M}_{\bm Y},A_A(u,\Omega)]&=-i\Delta_A(Y;A;u,\Omega)=-i\slashed\delta_{\bm Y}A_A,
\end{align}\es
which show that the operators $\mathcal{T}_f$ and $\mathcal{M}_{\bm Y}$ are supertranslation and superrotation generators, respectively.   Similar commutators between supertranslation/superrotation generators and shear tensor can be found in \cite{Barnich:2009se,Barnich:2011mi,He:2014laa,Kapec:2014opa,Strominger:2017zoo,Donnay:2022hkf}. However, there are several differences between our commutators and theirs. In our case, the operators $\mathcal T_f/\mathcal M_{\bm Y}$ are only made up of the radiative energy/angular momentum fluxes while the $Q_f/Q_Y$ in \cite{Strominger:2017zoo} is a combination of both the hard flux and the soft charge. Note that the test functions $f$ and vectors $\bm Y$ are also extended compared with those papers.   

\subsection{Intertwined Carrollian diffeomorphism}
So far, the discussion is completely the same as the one  in four dimensions. We expect a similar helicity flux operator in higher dimensions and will derive the intertwined Carrollian diffeomorphism by including this operator. The calculation is straightforward and we just show  the final result as follows 
\bs\label{commutators}
\begin{align}
    [\mathcal{T}_{f_1},\mathcal{T}_{f_2}]&={\rm C}_T(f_1,f_2)+i\mathcal{T}_{f_1\dot{f}_2-f_2\dot{f}_1},\\
[\mathcal{T}_f,\mathcal{M}_{\bm Y}]&=-i\mathcal{T}_{\bm Y(f)},\\
[\mathcal{T}_f,\mathcal{O}_{\bm h}]&=0,\\
[\mathcal{M}_{\bm Y},\mathcal{M}_{\bm Z}]&=i\mathcal{M}_{[\bm Y,\bm Z]}+i\mathcal{O}_{\bm o(\bm Y,\bm Z)},\\
[\mathcal{M}_{\bm Y},\mathcal{O}_{\bm h}]&=i\mathcal{O}_{\bm g(\bm Y,\bm h)},\\
[\mathcal{O}_{\bm h_1},\mathcal{O}_{\bm h_2}]&=-i\mathcal{O}_{[\bm h_1,\bm h_2]}.
\end{align}\es
In this closed algebra, three functions $f,f_1,f_2$ are smooth scalar fields on $\mathcal{I}^+$, while $\bm Y, \bm Z$ are smooth vector fields and $\bm o,\bm g,\bm h,\bm h_1,\bm h_2$ are 2-forms  on $S^{m}$. More explicitly, the components of the field $\bm h=\frac{1}{2}h_{AB}d\theta^A\wedge d\theta^B$ form an arbitrary  skew-symmetric matrix
\bea 
h_{AB}=-h_{BA}.
\eea 
The 2-form field $\bm o$ reads $\bm o=\frac{1}{2}o_{AB}d\theta^A\wedge d\theta^B$ with 
\bea 
o_{AB}(\bm Y,\bm Z)=\frac{1}{4}\left(\Theta_{AC}(\bm Y)\Theta^{C}_{\ B}(\bm Z)-\Theta_{AC}(\bm Z)\Theta^{C}_{\ B}(\bm Y)\right),\label{oAB}
\eea where $\Theta_{AB}(\bm Y)$ is a symmetric traceless tensor constructed by the vector field $\bm Y$
\bea 
\Theta_{AB}(\bm Y)=\nabla_AY_B+\nabla_BY_A-\frac{2}{m}\gamma_{AB}\nabla_CY^C.
\eea The third term of $\Theta_{AB}(\bm Y)$ has no contribution in \eqref{oAB}. Therefore, the field $\bm o$ is still independent of the dimension $d$. The 2-form field $\bm g=\frac{1}{2}g_{AB}d\theta^A\wedge d\theta^B$ is defined as
\bea 
g_{AB}(\bm Y,\bm h)&=&Y^C\nabla_C h_{AB}+\frac{1}{2}h_{A}^{\ C}(\nabla_BY_C-\nabla_CY_B)-\frac{1}{2}h_B^{\ C}(\nabla_AY_C-\nabla_CY_A)\nn\\&\equiv&Y^C\nabla_Ch_{AB}-\frac{1}{2}[\bm h,d\bm Y]_{AB},
\eea 
where we have defined a 2-form\footnote{Strictly speaking, $\bm Y$ is a vector field. One should map it to its dual vector field and then  use the exterior derivative operator to obtain the 2-form field.} 
\bea 
d\bm Y=\frac{1}{2}(\nabla_AY_B-\nabla_BY_A)d\theta^A\wedge d\theta^B.
\eea
The bracket between two skew-symmetric matrices $\left(h_1\right)_{AB}$ and $\left(h_2\right)_{AB}$ is still skew-symmetric
\bea 
[\bm h_1,\bm h_2]_{AB}=(h_1)_A^{\ C}(h_2)_{CB}-(h_2)_A^{\ C}(h_1)_{CB}.
\eea 
 Note that this definition for brackets between 2-forms is the same as the above $[\bm h,d\bm Y]$.

The central charge
\bea 
{\rm C}_T(f_1,f_2)=-\frac{im}{48\pi}\delta^{(m)}(0)\mathcal{I}_{f_1\dddot{f}_{\hspace{-2pt}2}-f_2\dddot{f}_{\hspace{-2pt}1}}
\eea reduces to the result of \cite{Liu:2023qtr} in four dimensions. The Dirac delta function $\delta^{(m)}(0)$ may be regularized using the zeta function or heat kernel method \cite{Li:2023xrr}. Interestingly, the factor $m$ is exactly the number of independent transverse propagating  degrees of freedom for  massless vector field in $d$ dimensions. This agrees with the conclusion that the central charge is proportional to the number of propagating degrees of freedom \cite{Liu:2024nkc}. The closed algebra generates the intertwined Carrollian diffeomorphism in which the new operator $\mathcal{O}_{\bm h}$ is defined as 
\bea 
\mathcal{O}_{\bm h}=\int du d\Omega h_{AB}(\Omega):\dot{A}^B A^A:.\label{helicityfluxoperator}
\eea This is the higher dimensional electromagnetic helicity flux operator which will  be discussed later. The structure of the Lie  algebra \eqref{commutators} is similar to the one in four dimensions except that the commutators between two helicity flux operators are non-vanishing which shows the non-Abelian property of the helicity flux operator in higher dimensions.

\paragraph{Ambiguities.}
In the expression of the superrotation generator, 
\be 
\mathcal{M}_{\bm Y}=\int du d\Omega \dot A^A(u,\Omega)\Delta_A(Y;A;u,\Omega),
\ee we can always separate the contribution from the helicity flux operator (seeing \eqref{sr})
\bea 
\mathcal{M}_{\bm Y}&=&\int du d\Omega \dot{A}^A(Y^C\nabla_CA_A+\frac{1}{2}\nabla_CY^C A_A)+\frac{1}{2}\int du d\Omega \dot{A}^A (\nabla_AY_C-\nabla_CY_A)A^C\nn\\&=&\int du d\Omega \dot{A}^A(Y^C\nabla_CA_A+\frac{1}{2}\nabla_CY^C A_A)-\frac{1}{2}\mathcal{O}_{\bm h=d\bm Y},
\eea where the first part is exactly the same as the superrotation generator by treating $A_A$ as a scalar field while the second part is the helicity flux operator.
We may define a one-parameter family of the operator 
\begin{align}
\mathcal{M}^{(\lambda)}_{\bm Y}=\mathcal{M}_{\bm Y}+\lambda \mathcal{O}_{\bm h=d\bm Y}\label{MYlambda}
\end{align}
such that 
\bea 
\mathcal{M}^{(1/2)}_{\bm Y}=\int du d\Omega \dot{A}^A(Y^C\nabla_AA_A+\frac{1}{2}\nabla_CY^CA_A).
\eea 
The commutators become
\bs \label{comlambda}
\begin{align}
[\mathcal{T}_{f_1},\mathcal{T}_{f_2}]&={\rm C}_T(f_1,f_2)+i\mathcal{T}_{f_1\dot{f_2}-f_2\dot{f}_1},\\
[\mathcal{T}_f,\mathcal{M}_{\bm Y}^{(\lambda)}]&=-i\mathcal{T}_{\bm Y(f)},\\
[\mathcal{T}_f,\mathcal{O}_{\bm h}]&=0,\\
[\mathcal{M}_{\bm Y}^{(\lambda)},\mathcal{M}_{\bm Z}^{(\lambda)}]&=i\mathcal{M}^{(\lambda)}_{[\bm Y,\bm Z]}+i\mathcal{O}_{\bm o^{(\lambda)}(\bm Y,\bm Z)},\\
[\mathcal{M}_{\bm Y}^{(\lambda)},\mathcal{O}_{\bm h}]&=i\mathcal{O}_{\bm g^{(\lambda)}(\bm Y,\bm h)},\\
[\mathcal{O}_{\bm h_1},\mathcal{O}_{\bm h_2}]&=-i\mathcal{O}_{[\bm h_1,\bm h_2]},
\end{align} \es where 
\begin{align}
    \bm g^{(\lambda)}(\bm Y,\bm h)&=\bm g(\bm Y,\bm h)+\lambda[\bm h,d\bm Y],\\
     \bm o^{(\lambda)}(\bm Y,\bm Z)&=\bm o(\bm Y,\bm Z)-\lambda d[\bm Y,\bm Z]+\lambda \bm g(\bm Y,d\bm Z)-\lambda\bm g(\bm Z,d\bm Y)-\lambda^2[d\bm Y,d\bm Z].\label{olambda}
\end{align} For general $\lambda$, there is no simplification. However,
when $\lambda=\frac{1}{2}$, we find\footnote{For more technical details, please consult Appendix \ref{comappendix}.}
\bs\begin{align}
    \left(\bm g^{(1/2)}(\bm Y,\bm h)\right)_{AB}&=Y^C\nabla_C h_{AB},\\
    \left(\bm o^{(1/2)}(\bm Y,\bm Z)\right)_{AB}&=-R_{ABCD}Y^CZ^D,\label{o12}
\end{align}\es where $R_{ABCD}$ is the Riemann curvature tensor of the unit sphere $S^{m}$.

\paragraph{Several comments.} 
\begin{enumerate}
    \item Once we add the helicity flux operator, we can always deform the superrotation generator to \eqref{MYlambda} for any constant $\lambda$. The commutators \eqref{comlambda} are involved for general $\lambda$, though they are equivalent to the former ones. 
    We can compute the infinitesimal transformation of the fundamental field 
    \be 
    \Delta^{(\lambda)}_A(Y;A;u,\Omega)=i[\mathcal{M}_{\bm Y}^{(\lambda)},A_A(u,\Omega)]
    \ee with\footnote{See \eqref{superduality} for the definition of $\Delta_A(d\bm Y;A;u,\Omega)$.}
    \bea 
     \Delta^{(\lambda)}_A(Y;A;u,\Omega)&=&\Delta_A(Y;A;u,\Omega)+\lambda \Delta_A(d\bm Y;A;u,\Omega)\nn\\&=&Y^C\nabla_CA_A+\frac{1}{2}\nabla_CY^C A_A+(\frac{1}{2}-\lambda)(\nabla_AY_C-\nabla_CY_A)A^C.\label{infvA}
    \eea 
    There are two  candidates of $\lambda$ that make the commutators simpler.
    \begin{itemize}
        \item The first choice is $\lambda=0$ which corresponds to the original commutators \eqref{commutators}. In this case, $\mathcal{M}_{\bm Y}$ could form a closed subalgebra $so(1,d-1)$ for $\bm Y$ being the CKVs since $\bm o(\bm Y,\bm Z)$ would be 0.  Therefore, $\mathcal{M}_{\bm Y}$ forms a faithful representation of the Lorentz algebra and it is indeed the angular momentum flux operator. 
        
        \item The second choice is $\lambda=\frac{1}{2}$ such that the commutators \eqref{comlambda} and the infinitesimal variation \eqref{infvA} are much simpler. However, $\mathcal{M}^{(1/2)}_{\bm Y}$ does not form a closed subalgebra even for $\bm Y$ being the CKVs due to the anomalous term associated with the Riemann curvature tensor in \eqref{o12}. Therefore, it is not the usual angular momentum flux operator and the physical interpretation of $\mathcal{M}^{(1/2)}_{\bm Y}$ is much more obscured in this case. 
    \end{itemize}

    \item In the derivation of the commutators, we just used  the skew symmetry, interchangeable symmetry and Bianchi identities of the Riemann tensor. Therefore, the algebra is still true on a general Carrollian manifold 
    \be 
    \mathcal{N}=\mathbb{R}\times N
    \ee where $N$ any smooth Riemannian manifold. This has been checked in Appendix \ref{hyper} and we just need to replace the Riemann curvature tensor of $S^{m}$ to the one of $N$. The effect of the geometry of the Carrollian manifold becomes obvious for the choice of $\lambda=\frac{1}{2}$. We just list two cases. 
    \begin{itemize}
        \item Future null infinity with $N=S^{m}$. The  Riemann curvature tensor is 
        \be 
R_{ABCD}=\gamma_{AC}\gamma_{BD}-\gamma_{AD}\gamma_{BC}.
\ee 
     Then $\bm o^{(1/2)}(\bm Y,\bm Z)$ becomes 
     \bea 
     \left(\bm o^{(1/2)}(\bm Y,\bm Z)\right)_{AB}=Z_AY_B-Y_AZ_B.
     \eea 
     \item Rindler horizon\footnote{Rindler horizon is a Carrollian manifold with the null vector $\partial_u$ where $u$ is defined as $u=\tau-\log\rho$ in \cite{Unruh:1976db}. The coordinate $\tau$ and $\rho$ are related to the Cartesian coordinates via the relation 
     \be 
     t=\rho\sinh\tau,\quad x^{d-1}=\rho\cosh\tau.\nn
     \ee Note that $\tau$ is the Rindler time \cite{Rindler:1966zz} which is also the time of an accelerated observer.  
    One can also check that 
     \be 
     \partial_u=\partial_\tau\nn
     \ee which is equivalent to the Lorentz boost along $x^{d-1}$ direction. The Rindler time is periodic in the imaginary direction which indicates a thermal bath detected by the accelerated observer \cite{Unruh:1976db,Bisognano:1975ih,Bisognano:1976za}. As a consequence, the null coordinates $u$ is also periodic after Wick rotation. To understand the Unruh effect from the Carrollian perspective, one may consider the Carrollian correlators in Minkowski vacuum instead of the Rindler vacuum. The Carrollian correlators in Rindler vacuum have been explored in \cite{Li:2024kbo}. To work in the Minkowski vacuum, one should develop the method to study correlators in thermal Carrollian field theory. } with $N=\mathbb{R}^{m}$. The Riemann curvature tensor is always zero and 
     \be 
     \bm o^{(1/2)}(\bm Y,\bm Z)=0.
     \ee As a consequence, the helicity flux operator disappears in the commutator 
     \be 
     [\mathcal{M}^{(1/2)}_{\bm Y},\mathcal{M}^{(1/2)}_{\bm Z}]=i\mathcal{M}^{(1/2)}_{[\bm Y,\bm Z]}.
     \ee 
    \end{itemize}

  \item Actually, we can deform the superrotation generator further by 
    \be 
   \widetilde{ \mathcal{M}}_{\bm Y}=\mathcal{M}_{\bm Y}+\mathcal{O}_{\bm \tau},
    \ee where $\bm\tau$ is an unspecified 2-form field on $S^{m}$. Therefore, we find 
    \bea 
    \widetilde{\Delta}_A(\bm Y,\tau;A;u,\Omega)=\Delta_A(\bm Y;A;u,\Omega)-\tau_{AC}A^C.\label{withtorsion}\eea Note that $\bm\tau$ may also depend on $\bm Y$, therefore we may write it more explicitly as 
    \be 
    \bm\tau=\bm\tau_{\bm Y}.
    \ee The commutators are modified to 
    \bs\begin{align}
        [\widetilde{\mathcal{M}}_{\bm Y},\widetilde{\cal M}_{\bm Z}]&=i\widetilde{\cal M}_{[\bm Y,\bm Z]}+i\mathcal{O}_{\widetilde{\bm o}(\bm Y,\bm Z)}\label{srtilde},\\
        [\widetilde{\cal M}_{\bm Y},\mathcal{O}_{\bm h}]&=i\mathcal{O}_{\widetilde{\bm g}(\bm Y,\bm h)}
    \end{align}\es while other commutators remain the same. The 2-form fields are 
    \bs\begin{align}
        \widetilde{\bm o}(\bm Y,\bm Z)&=\bm o(\bm Y,\bm Z)+\bm g(\bm Y,\bm \tau_{\bm Z})-\bm g(\bm Z,\bm\tau_{\bm Y})-\bm\tau_{[\bm Y,\bm Z]}-[\bm \tau_{\bm Y},\bm\tau_{\bm Z}],\\
        \widetilde{\bm g}(\bm Y,\bm h)&=\bm g(\bm Y,\bm h)-[\bm \tau_{\bm Y},\bm h].
    \end{align}\es 
    To discard the  operator $\mathcal{O}$ in the commutator \eqref{srtilde}, we may impose the condition
    \be 
    \widetilde{\bm o}(\bm Y,\bm Z)=0.
    \ee 
    This is a set of non-linear equations in general dimensions. In four dimensions, the 2-form field is proportional to the Levi-Civita tensor of $S^2$ and the last term disappears. In this case, the equation becomes linear.  It is not likely that there are universal solutions for arbitrary smooth vectors $\bm Y$ and $\bm Z$ on the unit sphere in general dimensions. Therefore, it is impossible to avoid the helicity flux operator once we introduce the superrotation.
\end{enumerate}
 We will show that the ambiguities  are relevant with the choice of the connection when defining covariant variation in section \ref{sec3.3} in the framework of bulk reduction.

\subsection{Gauge transformation}
In the previous subsections, we have obtained three independent operators $\{\mathcal{T}_f,\mathcal{M}_{\bm Y},\mathcal{O}_{\bm h}\}$ which are expected to be physical observables. For a massless theory with a non-zero spin, a necessary condition is that any physical observable should be gauge invariant. In this subsection, we will deal with this issue at the boundary. 

The Carrollian manifold $\mathcal{I}^+$ is  $d-1$ dimensional and the gauge parameter $\epsilon$ associated with the vector field $A_A$ may depend on all the coordinates of the manifold 
\be 
\epsilon=\epsilon(u,\Omega).\label{gauge}
\ee The gauge transformation for the fundamental field $A_A(u,\Omega)$ is assumed to be 
\be 
A_A\to A_A+\partial_A\epsilon.
\ee However, if $\epsilon$ is any smooth field on $\mathcal{I}^+$, as \eqref{gauge}, one can always use this gauge transformation to reduce the number of the degrees of freedom by 1. To clarify this point, we set $d=4$ and then $A_A$ has two independent components. We decompose it into two scalar functions
\be A_A=\partial_A\Phi+\epsilon_A^{\ B}\partial_B\Psi,\label{divideA}
\ee and choose 
\be \epsilon=-\Phi
\ee such that $A_A$ is transformed to $A'_A=\epsilon_{A}^{\ B}\partial_B\Psi$. This is unacceptable since we assumed that the number of independent degrees  of freedom of $A_A$ is $d-2=2$. The argument can be extended to general $d$ dimensions. Therefore, we conclude that $\epsilon$ cannot be unconstrained. Now we use the gauge invariance of the operator $\mathcal{T}_f$ to impose constraints on $\epsilon$. Under a general  transformation \eqref{gauge}, we have 
\bea 
\mathcal{T}_f\to\mathcal{T}_f+\int du d\Omega f(u,\Omega)(2\dot{A}^A\partial_A\dot\epsilon+\partial^A\dot\epsilon \partial_A\dot\epsilon).
\eea It is invariant only for 
\be 
\dot\epsilon=0\quad\Rightarrow\quad \epsilon=\epsilon(\Omega).
\ee Therefore, the gauge invariance of $\mathcal{T}_f$ suggests the following gauge transformation of the vector field 
\be 
A_A(u,\Omega)\to A_A(u,\Omega)+\partial_A\epsilon(\Omega).\label{gauge2}
\ee Now we can check the gauge invariance of the operator $\mathcal{M}_{\bm Y}$ and $\mathcal{O}_{\bm h}$  as follows
\bs
\begin{align}
  \delta_\epsilon\mathcal{M}_{\bm Y}&= \int du d\Omega \dot{A}^A (Y^C\nabla_C\nabla_A\epsilon+\frac{1}{2}\nabla_CY^C \nabla_A\epsilon+\frac{1}{2}(\nabla_AY_C-\nabla_CY_A)\nabla^C\epsilon)\overset{\text{?}}{=}0,\\ 
  \delta_\epsilon \mathcal{O}_{\bm h}&= \int du d\Omega \dot A^A h_{BA}\partial^B\epsilon\overset{\text{?}}{=}0.
\end{align}

\es 
The symbol $\overset{\text{?}}=0$ indicates that the operators  $\mathcal M_{\bm Y}$ and $\mathcal O_{\bm h}$ are not necessary to be gauge invariant. One should impose further conditions on the configuration $A_A(u,\Omega)$, the parameters $\bm Y,\ \bm h$ and/or the gauge parameter $\epsilon$. There are three ways to solve the problem.  
\begin{enumerate}
    \item We constrain the gauge parameter  $\epsilon(\Omega)$ to be a constant 
    \be 
    \epsilon(\Omega)=\text{const.}\label{epcon}
    \ee independent of the angular coordinates. Then there is no large gauge transformation and the gauge invariance of the flux operators is preserved.
    \item We may assume the parameters $\bm Y,\ \bm h$ are time-independent. This is consistent with  the absence of the non-local terms in the commutators between the flux operators. Then the gauge invariance of $\mathcal{M}_{\bm Y}$ and $\mathcal{O}_{\bm h}$ can be saved by imposing the condition 
    \be 
    A_A(u=\infty,\Omega)-A_A(u=-\infty,\Omega)=0.\label{Aacondition}
    \ee In this case, the gauge parameter can be any smooth vector $\epsilon(\Omega)$ on the sphere. However, the condition \eqref{Aacondition} indicates that the permanent change of $A_A(u,\Omega)$ during the radiative process should be vanishing. This condition is too strong and rules out many interesting physical phenomena. 
    \item We still assume the parameters $\bm Y,\ \bm h$ are time-independent but relax the condition \eqref{Aacondition}, namely allow
    \be 
     A_A(u=\infty,\Omega)-A_A(u=-\infty,\Omega)\not=0.
    \ee  In this case, the variation of the flux operators are not invariant for general $\epsilon(\Omega)$. However, it is possible that the transformation generated by $\epsilon(\Omega)$ becomes a large gauge transformation. There is no problem that the fluxes $\mathcal{M}_{\bm Y}$ and $\mathcal{O}_{\bm h}$ transform under large gauge transformations.
\end{enumerate} 
We will return to this point in the following section and Appendix \ref{gaugechoice}.

\section{Bulk reduction}\label{bulk}
In the previous discussion, we elaborated on the boundary theory on a Carrollian manifold by assuming that the symplectic form and the transformation law of the boundary field are formally the same as those in four dimensions. In this section, we will confirm these assumptions using the method of bulk reduction. 

\subsection{Equation of motion and symplectic form}
The starting point is to embed the co-dimension one Carrollian manifold into  Minkowski spacetime in which the action of the electromagnetic field is 
\bea 
S[a]=-\frac{1}{4}\int d^dx f_{\mu\nu}f^{\mu\nu},\quad f_{\mu\nu}=\partial_\mu a_\nu-\partial_\nu a_\mu.\label{actionvector}
\eea
The action is invariant under the gauge transformation 
\be 
a_\mu\to a_\mu+\partial_\mu\epsilon_{\text{bulk}},
\ee where $\epsilon_{\text{bulk}}$ is a local function of spacetime coordinates. We will impose the following fall-off condition for the vector field\footnote{We usually omit the superscript $(0)$ in the leading coefficients to simplify notation.}
\bea 
a_\mu(x)=\frac{A_\mu(u,\Omega)}{r^{\Delta}}+\sum_{k=1}^\infty \frac{A_\mu^{(k)}(u,\Omega)}{r^{\Delta+k}}\label{falloffamu0}
\eea  with 
\bea 
\Delta=\frac{d-2}{2}.
\eea The fall-off condition is the same as the scalar field \cite{Li:2023xrr}.  A rigorous treatment on the fall-off condition can be found in \cite{Satishchandran:2019pyc}. See also the references \cite{Kapec:2014zla,Campoleoni:2017qot,He:2019jjk,He:2019pll}.
 Notice that there are two branches from solving the equation of motion. The radiative modes have the same fall-off behaviour as ours and the Coulombic order starts at $\mathcal{O}(r^{3-d})$.  In odd dimensions, the expansion power of the radiative modes have half integer and then one should include both the integer and half integer fall-offs in general \cite{Campoleoni:2017qot}. In our work, we only consider the radiative modes.
To transform to the retarded coordinate system, we define the following null vectors 
\bea 
n^\mu=(1,n^i),\quad \bar{n}^\mu=(-1,n^i),
\eea where $n^i$ is the normal vector of $S^{m}$ 
\bea 
n^i=\frac{x^i}{r}.
\eea In terms of these two null vectors, we can define  timelike and spacelike vectors 
\bea 
\bar{m}^\mu=\frac{1}{2}(n^\mu-\bar{n}^\mu),\quad m^\mu=\frac{1}{2}(n^\mu+\bar n^\mu)
\eea 
such that the Cartesian coordinates and the retarded coordinates are related by 
\bea 
x^\mu=u\bar{m}^\mu+r n^\mu.
\eea Therefore, the Jacobi matrix takes the form
\bea 
\frac{\partial x^\mu}{\partial x^\alpha}=\bar{m}^\mu\delta_\alpha^u+n^\mu\delta_\alpha^r-r Y^\mu_A\delta^A_\alpha
\eea where 
\bea 
Y^\mu_A=-\nabla_An^\mu=-\nabla_A \bar n^\mu=-\nabla_A m^\mu.
\eea 

The components of the  vector potential in retarded coordinates are 
\bs\label{falloffretarded}
\begin{align}
    a_A(x)&=\frac{A_A(u,\Omega)}{r^{\Delta-1}}+\sum_{k=1}^\infty \frac{A_A^{(k)}(u,\Omega)}{r^{\Delta+k-1}},\\
    a_{u}(x)&=\frac{A_u(u,\Omega)}{r^\Delta}+\sum_{k=1}^\infty\frac{ A_u^{(k)}(u,\Omega)}{r^{\Delta+k}},\\
    a_r(x)&=\frac{  A_r(u,\Omega)}{r^\Delta}+\sum_{k=1}^\infty\frac{ A_r^{(k)}(u,\Omega)}{r^{\Delta+k}}
\end{align}
\es with 
\bea 
&&A_A^{(k)}=-Y^\mu_A A_\mu^{(k)},\quad A_u^{(k)}=\bar{ m}^\mu A_\mu^{(k)},\quad  A_r^{(k)}=n^\mu A_\mu^{(k)},\quad k=0,1,2,\cdots.
\eea The $k=0$ components are the leading coefficients in the asymptotic expansion.
Interestingly, these expressions can be unified as 
\bea 
 { A_\alpha^{(k)}=\bar{N}_\alpha^{\ \mu}A^{(k)}_\mu,\quad \bar{N}_\alpha^{\ \mu}=\bar{m}^\mu\delta_\alpha^u+n^\mu\delta^r_\alpha-Y^\mu_A\delta_\alpha^A.}
\eea Inversely, this is 
\bea 
A_\mu^{(k)}=N_\mu^{\ \alpha}A^{(k)}_\alpha,\quad N_\mu^{\ \alpha}=-n_\mu\delta^\alpha_u+m_\mu\delta^\alpha_r-Y_{\mu}^{A}\delta^\alpha_A.\label{Nd}
\eea The tensors $N_\mu^{\ \alpha}$ and $\bar{N}_\alpha^{\ \mu}$  have appeared in \cite{Liu:2023jnc} and they are used to transform components between Cartesian and retarded coordinates. 

The fall-off conditions \eqref{falloffretarded} impose constraints on the gauge parameter $\epsilon_{\text{bulk}}$. We expand the parameter near $\mathcal{I}^+$ as 
\be 
\epsilon_{\text{bulk}}={ \epsilon_0}+\sum_{k=0}^\infty\frac{ \epsilon^{(k)}_{\text{bdy}}(u,\Omega)}{r^{\widetilde{\Delta}+k}},\label{fallepsilon}
\ee where $\widetilde{\Delta}$ has not been fixed at this moment.  Then $a_A$ transforms as 
\bea 
\delta_{\epsilon_{\text{bulk}}} a_A=\partial_A\epsilon_0+\sum_{k=0}^\infty \frac{\partial_A\epsilon^{(k)}_{ \text{bdy}}(u,\Omega)}{r^{\widetilde{\Delta}+k}}.
\eea To preserve the fall-off conditions \eqref{falloffretarded}, the constant $\widetilde{\Delta}$ should be 
\be 
\widetilde{\Delta}=\Delta-1
\ee and the term $\epsilon_0$ should be angle-independent 
\be 
\partial_A\epsilon_0=0.\label{anin}
\ee Then the transformation of the fundamental field $A_A$ is
\be 
\delta_{\epsilon_{\text{bdy}}}A_A=\partial_A\epsilon_{\text{bdy}}
\ee 
To preserve the fall-off condition of $a_u$, we find 
\be 
\dot{\epsilon}_{\text{bdy}}=0\quad\Rightarrow\quad \epsilon_{\text{bdy}}=\epsilon(\Omega)\label{angleau}
\ee and 
\be 
\dot\epsilon_0=0\quad\Rightarrow\quad \epsilon_0=\text{const.}\label{epsilon0}
\ee When deriving the above condition, we have used the angle-independent condition \eqref{anin}. Note that $\epsilon_0=\text{const.}$  leads to a finite electric charge. The equation \eqref{angleau} is exactly the gauge transformation of the boundary fundamental field \eqref{gauge2} which is claimed by imposing the gauge invariance of the Hamiltonians. From the bulk point of view, it may lead to the residual gauge transformation that preserves the gauge fixing condition and the fall-off conditions. In general, it would become large gauge transformation once the corresponding charge is nontrivial. In Appendix \ref{gaugechoice}, we discuss the potential large gauge transformation and its consequences, though these parts are not closely related to the topic of this paper.

The partial derivative is expressed in terms of the derivatives of retarded coordinates
\bea 
\partial_\mu=-n_\mu \partial_u+m_\mu\partial_r-\frac{1}{r}Y^A_\mu\partial_A.
\eea Therefore, the electromagnetic field is 
\bea 
f_{\mu\nu}&=&\sum_{k=0}^\infty \frac{f_{\mu\nu}^{(k)}}{r^{\Delta+k}}
\eea where 
\bea 
f_{\mu\nu}^{(k)}&=&(n_\nu N_\mu^{\ \alpha}-n_\mu N_\nu^{\ \alpha})\dot{A}_{\alpha}^{(k)}+(\Delta+k-1)(m_\nu N_\mu^{\ \alpha}-m_\mu N_\nu^{\ \alpha})A_\alpha^{(k-1)}\nn\\&&-Y^A_\mu \nabla_A(N_\nu^{\ \alpha} A_\alpha^{(k-1)})+Y^A_\nu\nabla_A(N_\mu^{\ \alpha}A_\alpha^{(k-1)})\nn\\&=&A_{\mu\nu}^{\ \ \alpha}\dot{A}_{\alpha}^{(k)}+(\Delta+k-1)B_{\mu\nu}^{\ \ \alpha}A_\alpha^{(k-1)}+C_{\mu\nu}^{\ \ \alpha}A_\alpha^{(k-1)}+D_{\mu\nu}^{\ \ \alpha A}\nabla_A A_\alpha^{(k-1)},\quad k=0,1,2,\cdots.\nn\\
\eea We use the convention 
\bea 
A^{(-n)}_\alpha=0,\quad n=1,2,\cdots
\eea and the tensors $A_{\mu\nu}^{\ \ \alpha},\ B_{\mu\nu}^{\ \ \alpha},\ C_{\mu\nu}^{\ \ \alpha},\ D_{\mu\nu}^{\ \ \alpha A}$ are defined in \eqref{abcd}. The first few orders of $f_{\mu\nu}^{(k)}$ are 
\bs\begin{align}
    f_{\mu\nu}^{(0)}&=A_{\mu\nu}^{\ \ \alpha}\dot{A}_\alpha,\\
    f_{\mu\nu}^{(1)}&=A_{\mu\nu}^{\ \ \alpha}\dot{A}^{(1)}_\alpha+\Delta B_{\mu\nu}^{\ \ \alpha}A_\alpha+C_{\mu\nu}^{\ \ \alpha}A_\alpha+D_{\mu\nu}^{\ \ \alpha A}\nabla_A A_\alpha.
\end{align}\es 

The equation of motion in the bulk may be solved order by order 
\bea 
\partial_\mu f^{\mu\nu}=0\Rightarrow
n_\mu \dot{f}^{\mu\nu(k)}+(\Delta+k-1)m_\mu f^{\mu\nu(k-1)}+Y^A_\mu \nabla_A f^{\mu\nu(k-1)}=0,\quad k=0,1,2,\cdots.
\eea 
The equation of motion can be expanded in the basis $n^\nu, m^\nu, Y^{\nu A}$ and then we find 
\bs\begin{align}
   & \ddot{A}_r^{(k)}+(\Delta+k-1)\dot{A}_u^{(k-1)}+(\Delta+k-1-m)(\dot{A}_r^{(k-1)}+(\Delta+k-2)A_u^{(k-2)})-\nabla^A\dot{A}_A^{(k-1)}+\nabla^2 A_u^{(k-2)}=0,\label{nnu}\\
   &(\Delta+k-1-m)(\dot{A}_r^{(k-1)}+(\Delta+k-2)A_u^{(k-2)})+(\Delta+k-3)\nabla^AA_A^{(k-2)}+\nabla^2A_r^{(k-2)}=0,\label{mnu}\\
   &(2k-2)\dot{A}_C^{(k-1)}+\nabla_C\dot{A}_r^{(k-1)}+(\Delta+k-m)((\Delta+k-3)A_C^{(k-2)}-\nabla_C A_u^{(k-2)}+\nabla_CA_r^{(k-2)})\nn\\&+\nabla^A(\nabla_AA_C^{(k-2)}-\nabla_CA_A^{(k-2)})=0.
\end{align}\es 

For $k=0$, we only find 
\bea 
\ddot{A}_r=0.
\eea For $k=1$, we find 
\bs\begin{align}
&\ddot{A}_r^{(1)}+\Delta \dot{A}_u+(\Delta-m)\dot{A}_r-\nabla^A\dot{A}_A=0,\label{Aueom}\\
&\dot{A}_r=0,\label{dotAr0}\\
& \nabla_C\dot{A}_r=0.
\end{align}\es 
Note that the leading coefficient $A_r=\varphi(\Omega)$ is independent of the retarded time. 

The symplectic form at $\mathcal{I}^+$ can be obtained by taking the limit 
\bea 
\lim{}\!_{+}=\lim_{r\to\infty,\ u\ \text{fixed}}
\eea for the symplectic form on a $r=\text{const.}$ surface $\mathcal{H}_r$
\bea 
\bm\Omega(\delta A;\delta A;A)&=&-\lim{}\!_{+}\int_{\mathcal{H}_r}\left(d^{d-1}x\right)_\mu \delta f^{\mu}_{\ \nu}\wedge \delta a^\nu \nn\\&=&\int du d\Omega (n_\nu N_\mu^{\ \alpha}-n_\mu N_\nu^{\ \alpha})\delta \dot{A}_\alpha\wedge { N^{\nu\beta}}\delta A_\beta m^\mu\nn\\&=&\int du d\Omega \delta A^A\wedge \delta\dot{A}_A,\label{symplectic}
\eea where we have used the identities in Appendix \ref{identities}. In the last step, we also used the equation of motion to set $\dot{A}_r=0$. The symplectic form \eqref{symplectic} is exactly the same as we have assumed in \eqref{symp1}. The derivation is independent of the gauge choice. In the gauge $a_r=0$ we find 
\bea 
\Delta A_u=\nabla^A A_A+\widetilde{\varphi}(\Omega)\label{Au}
\eea by solving the equation of motion \eqref{Aueom} and $\widetilde{\varphi}(\Omega)$ is an integration constant.

\subsection{Canonical quantization}
We can also derive the commutator \eqref{comAA} within the framework of bulk reduction. In the Lorenz gauge $\partial_\mu a^\mu=0$, the vector field $a_\mu(t,\bm x)$ can be expanded as the superposition of positive and negative frequency modes 
\begin{align}
a_\mu(t,\bm x)&=\sum_{a}\int \frac{d^{d-1}\bm k}{\sqrt{(2\pi)^{d-1}}}\frac{1}{\sqrt{2\omega_{\bm{k}}}}(e^{-i\omega t+i\bm{k}\cdot\bm{x}}\epsilon_\mu^{a*}(\bm k) b_{a,\bm{k}}+e^{i\omega t-i\bm{k}\cdot\bm{x}}\epsilon_\mu^a(\bm k) b_{a,\bm{k}}^\dagger),\label{modeexpMink}
\end{align}
where $\epsilon_\mu^a(\bm k)$ are the polarization vectors and they satisfy the orthogonality and completeness relations
\bs \label{polarization}
\begin{align}
\sum_{a,b}\epsilon^{a *}_\mu \delta_{ab}\epsilon_\nu^b&=\gamma_{\mu\nu},\\
\sum_{\mu,\nu}\epsilon^{a *}_\mu \gamma^{\mu\nu}\epsilon_\nu^b&=\delta^{ab},
\end{align} \es where the symmetric tensor $\gamma_{\mu\nu}$ is 
\begin{align}
\gamma_{\mu\nu}=\eta_{\mu\nu}-\frac{1}{2}(n_\mu(\bm k)\bar n_\nu(\bm k)+n_\nu(\bm k)\bar n_\mu(\bm k))=Y_\mu^A Y_\nu^B\gamma_{AB}.
\end{align} 
The label $a=1,2,\cdots,m$ denotes the $m$ independent transverse modes. A convenient representation of the polarization vectors would be 
\bea 
\epsilon_\mu^a=Y^A_\mu e^a_A
\eea with $e^a_A$ the vielbeins of the unit sphere $S^{m}$ that satisfy the orthogonality and completeness relations
\bea 
e^a_A e^b_B\gamma^{AB}=\delta^{ab},\qquad \gamma_{AB}=e^a_A e^b_B \delta_{ab}.
\eea Therefore, the constraints \eqref{polarization} for the polarization vectors are satisfied automatically. 

The annihilation and creation operators $b_{a,\bm k}$ and $b^\dagger_{a,\bm k}$ satisfy the standard commutation relations
\bea 
[b_{a,\bm k}, b_{b,\bm k'}]=[b^\dagger_{a,\bm k}, b^\dagger_{b,\bm k'}]=0,\qquad [b_{a,\bm k}, b^\dagger_{b,\bm k'}]= \delta^{(d-1)}(\bm k-\bm k'){ \delta_{ab}}.
\eea The vacuum state $|0\rangle$ is annihilated by the operator $b_{a,\bm k}$
\be 
b_{a,\bm k}|0\rangle=0.
\ee 
Expanding the plane wave into  a superposition of spherical waves\footnote{The spherical harmonic function $Y_{\bm \ell}(\Omega)$ in higher dimensions can be found in \cite{Higuchi:1986wu} and has been reviewed in \cite{Li:2023xrr}.}
\begin{align}
e^{i \bm{k}\cdot \bm x}=\frac{2(d-3)!!\pi^{(d-1)/2}}{\Gamma((d-1)/2)}\sum_{\bm\ell}i^{\ell_{m}} j^{d-1}_{\ell_{m}}(\omega r)Y^*_{\bm\ell}(\Omega_k)Y_{\bm\ell}(\Omega),
\end{align} we may find the following asymptotic expansion near $\mathcal{I}^+$
\begin{align}
    a_\mu(t,\bm x)=\frac{A_\mu(u,\Omega)}{r^{\Delta}}+\cdots, \label{falloffamu}
\end{align} where the leading term is 
\begin{align}
    A_\mu(u,\Omega)&=\sum_{\bm\ell}\int_0^\infty \frac{d\omega}{\sqrt{4\pi\omega}}[c_{\mu;\omega,\bm\ell}e^{-i\omega u}Y_{\bm \ell}(\Omega)+c_{\mu;\omega,\bm \ell}^\dagger e^{i\omega u}Y^*_{\bm \ell}(\Omega)]
\end{align}
with the coefficients 
\begin{align} 
c_{\mu;\omega,\bm\ell}&=\omega^{m/2}e^{-i\pi m/4}\int d\Omega_k b_{a,\bm k}Y^*_{\bm \ell}(\Omega_k)\epsilon_\mu^{a *}(\bm k),\\
c_{\mu;\omega,\bm\ell}^\dagger&=\omega^{m/2}e^{i\pi m/4}\int d\Omega_k b^\dagger_{a,\bm k}Y_{\bm \ell}(\Omega_k)\epsilon_\mu^{a}(\bm k).
\end{align} 
The asymptotic expansion \eqref{falloffamu} is consistent with \eqref{falloffamu0}.  
Therefore, the transverse modes are 
\begin{align}
A_A(u,\Omega)=-Y_A^\mu A_\mu(u,\Omega)={ -}Y^\mu_A(\Omega)\sum_{\bm\ell}\int_0^\infty \frac{d\omega}{\sqrt{4\pi\omega}}[c_{\mu;\omega,\bm\ell}e^{-i\omega u}Y_{\bm \ell}(\Omega)+c_{\mu;\omega,\bm \ell}^\dagger e^{i\omega u}Y^*_{\bm \ell}(\Omega)]\label{canonicalA}
\end{align} and the commutator \eqref{comAA} could be checked using canonical quantization.

\subsection{Transformation law}\label{sec3.3}
The transformation law of the boundary field $A_A$ under Carrollian diffeomorphism can also be found from bulk reduction. The supertranslation vector \eqref{stv} and superrotation vector \eqref{srv} can be extended into the bulk near $\mathcal{I}^+$. Their expression can be found in \cite{Colferai:2020rte}
\bs\begin{align}
     \bm \xi_f&=f\partial_u-\frac{1}{r}\nabla^Af\partial_A+\frac{1}{m}\nabla^2f\partial_r+\cdots,\\
    \bm \xi_{\bm Y}&=\frac{u}{m}\nabla_AY^A\partial_u+(Y^A-\frac{u\nabla^A\nabla_CY^C}{mr})\partial_A-(\frac{r}{m}\nabla_AY^A-\frac{u\nabla_A\nabla^A\nabla_CY^C}{m^2})\partial_r+\cdots.
\end{align}\es 
The variation of $A_A$ under supertranslation can be read out from the Lie derivative of $a_\alpha$ along $\bm\xi_f$\footnote{Here the Lie derivative is associated with the bulk manifold.}
\bea 
\delta_f A_A=\lim{}\!_+r^{\Delta-1}  \cl_{\bm\xi_f}a_A=f\dot{A}_A
\eea and we confirm the transformation law \eqref{st}. We can also calculate the Lie derivative of $a_\alpha$ along $\bm\xi_{\bm Y}$ 
\bea 
\delta_{\bm Y} A_A&=&\lim{}\!_+r^{\Delta-1}\cl_{\bm\xi_{\bm Y}}a_A\nn\\&=&\frac{u}{m}\nabla_CY^C\dot A_A+Y^C\nabla_CA_A+\nabla_AY^C A_C+\frac{m-2}{2m}\nabla_CY^C A_A-\frac{1}{m}\nabla_A\nabla_CY^C \varphi(\Omega).\nn\\
\eea As has been explained, we may define a covariant variation as 
\bea 
\slashed\delta_{\bm Y} A_A=\delta_{\bm Y} A_A-\Gamma_A^{\ C}A_C-\text{inhomogeneous term}-\text{supertranslation term}.
\eea 
The affine connection is assumed to be symmetric in its indices 
\bea \Gamma_{AB}=\Gamma_{BA}.\label{symcon}
\eea It can be found by the invariance of the boundary metric under superrotation 
\bea 
\slashed\delta_{\bm Y}\gamma_{AB}=\delta_{\bm Y}\gamma_{AB}-\Gamma_{A}^{\ C}\gamma_{CB}-\Gamma_B^{\ C}\gamma_{AC}=0\quad\Rightarrow\quad \Gamma_{AB}=\frac{1}{2}\Theta_{AB}(\bm Y).
\eea 
The inhomogeneous term which is independent of $A_A$ has been subtracted in the covariant variation. We have also removed the term which is related to supertranslation. Therefore, 
\bea 
\slashed\delta_{\bm Y} A_A=Y^C\nabla_CA_A+\nabla_AY^C A_C+\frac{m-2}{2m}\nabla_CY^C A_A-\frac{1}{2}\Theta_{AC}(Y)A^C=\Delta_A(Y;A;u,\Omega).
\eea The $d$-dependence disappears and we find the same form as \eqref{sr}.

The previous discussion assumes that the connection is symmetric. However, we may add a torsion term to the affine connection and define 
\be 
\widetilde{\Gamma}_{AB}=\Gamma_{AB}+\tau_{AB},
\ee where $\Gamma_{AB}$ is still symmetric while the torsion term is antisymmetric 
\be 
\tau_{AB}=-\tau_{BA}.
\ee Then we should find the following covariant variation 
\bea 
\widetilde{\slashed\delta}_{\bm Y} A_A&=&\delta_{\bm Y} A_A-\widetilde{\Gamma}_A^{\ C}A_C-\text{inhomogeneous term}-\text{supertranslation term}\nn\\&=&\Delta_{A}(Y;A;u,\Omega)-\tau_A^{\ C}A_C.
\eea Interestingly, this variation corresponds to the ambiguity in the definition of the superrotation variation \eqref{withtorsion}. Moreover, if we set 
\be 
\tau_{AB}=\lambda(\nabla_AY_B-\nabla_BY_A),
\ee we get the one-parameter family operators $\mathcal{M}_{\bm Y}^{(\lambda)}$. Therefore, we may admit that the ambiguity of the superrotation generator is in one-to-one correspondence with the numerous choices of the connection. 

\subsection{Fluxes from bulk to boundary}
Now we will construct the fluxes related to Poincar\'e invariance of Minkowski spacetime. For the spacetime translation, the corresponding conserved current is the stress tensor which is a quadratic form of the electromagnetic field
\bea 
T_{\mu\nu}=f_{\mu\rho}f_{\nu}^{\ \rho}-\frac{1}{4}\eta_{\mu\nu}f_{\rho\sigma}f^{\rho\sigma}.
\eea The Killing vector may be parameterized by 
\bea 
\bm\xi_c=c^\mu\partial_\mu
\eea with $c^\mu$ a constant vector. The energy and momentum fluxes that arrived at $\mathcal{I}^+$ are 
\bea 
Q_{\bm\xi_c}&=& { -}\lim{}\!_+ \int_{\mathcal{H}_r} (d^{d-1}x)_\mu T^{\mu}_{\ \nu}\xi_c^\nu\nn\\&=&c^\nu\lim{}\!_+r^{2\Delta}\int du d\Omega  m^\mu T_{\mu\nu}.
\eea 
The stress tensor may be expanded asymptotically as 
\bea 
T_{\mu\nu}=\sum_{k=0}^\infty \frac{T_{\mu\nu}^{(k)}}{r^{2\Delta+k}}
\eea with 
\bea 
T^{(k)}_{\mu\nu}&=&\sum_{j=0}^k[ f_{\mu\rho}^{(j)}f_{\nu}^{\ \rho (k-j)}-\frac{1}{4}\eta_{\mu\nu}f_{\rho\sigma}^{(j)}f^{\rho\sigma(k-j)}],
\eea whose explicit form is presented in \eqref{stressk}. We only need the leading and subleading orders in this work\footnote{For simplicity, we choose the gauge $a_r=0$. For the stress tensor without imposing the gauge condition, see Appendix \ref{identities}.}
\bs\begin{align}
    T_{\mu\nu}^{(0)}&=n_\mu n_\nu \dot{A}_A\dot{A}_B\gamma^{AB},\\
    T_{\mu\nu}^{(1)}&=2\gamma^{AB}n_\mu n_\nu \dot{A}_A\dot{A}_B^{(1)}+\Delta(\bm Y_\mu^An_\nu+Y_\nu^An_\mu)\dot{A}_AA_u\nn\\&\quad +(\Delta-1)(Y_\mu^AY_\nu^B+Y_\nu^AY_\mu^B-\gamma^{AB}(n_\mu m_\nu+n_\nu m_\mu))\dot{A}_AA_B\nn\\&\quad -2n_\mu n_\nu \dot{A}_A\nabla^A A_u+(\gamma^{BC}Y_\nu^An_\mu+\gamma^{BC}Y_\mu^An_\nu-\gamma^{AB}Y_\nu^Cn_\mu-\gamma^{AB}Y_\mu^Cn_\nu)\dot{A}_B\nabla_AA_C.
\end{align}\es 
Then, the energy and momentum fluxes are 
\bea 
Q_{\bm\xi_c}&=&c^\nu \int du d\Omega m^\mu T_{\mu\nu}^{(0)}=c^\nu \int du d\Omega n_\nu \dot{A}_A\dot{A}_B\gamma^{AB}.
\eea More explicitly, $Q_{\bm\xi_c}$ is the energy flux for $c^\mu=\delta^\mu_0$ and the momentum flux in $i$-th direction for $c^\mu=\delta^\mu_i$. The local operator 
\bea 
T(u,\Omega)=:\gamma^{AB}\dot{A}_A\dot{A}_B:
\eea is the energy flux density which is the radiative energy across $\mathcal{I}^+$ per unit time and unit solid angle
\bea 
\frac{dE}{du d\Omega}=-T(u,\Omega).
\eea In the Fourier space\footnote{This is actually a generalized Fourier transform since we also transform the angular directions at the same time. }, the operator $T(u,\Omega)$ is transformed to the supertranslation generator 
\bea 
\mathcal{T}_f=\int du d\Omega f(u,\Omega)T(u,\Omega), \label{tfdef}
\eea where the $T(u,\Omega)$ is normal ordered. 

For Lorentz transformation that is parameterized by the Killing vector 
\bea 
\bm\xi_\omega=\omega^{\mu\nu}(x_\mu\partial_\nu-x_\nu\partial_\mu),
\eea the angular momentum and center-of-mass fluxes are 
\bea 
Q_{\bm\xi_\omega}&=&{ -}\lim{}\!_+\int_{\mathcal{H}_r}(d^{d-1}x)_\mu T^{\mu}_{\ \nu}\xi^\nu_\omega\nn\\&=&2\omega^{\rho\nu}\lim{}\!_+ r^{2\Delta}\int du d\Omega x_\rho m^\mu T_{\mu\nu}\nn\\&=&2\omega^{\rho\nu}\int du d\Omega (u \bar{m}_\rho m^\mu T_{\mu\nu}^{(0)}+n_\rho m^\mu T_{\mu\nu}^{(1)})\nn\\&=&\frac{1}{2}\omega^{\mu\nu}\int du d\Omega u n_{\mu\nu} T(u,\Omega)-\omega^{\mu\nu}\int du d\Omega Y_{\mu\nu}^A(\Delta A_u  \dot{A}_A+\dot{A}^C \nabla_AA_C-\dot{A}^C\nabla_CA_A) \nn\\
&=&{-}\omega^{\mu\nu}\int du d\Omega \frac{1}{m}u \nabla_AY^A_{\mu\nu} T(u,\Omega)+\omega^{\mu\nu}\int du d\Omega  \dot{A}_C\nabla_AA_B(-\gamma^{AB}Y_{\mu\nu}^C-\gamma^{BC}Y_{\mu\nu}^A+\gamma^{CA}Y_{\mu\nu}^B).\nn\\
\eea To get this result, we have used the relation \eqref{Au} and  thrown out the total derivative terms. The first integral is a supertranslation  generator with $f=\frac{1}{m}u \omega^{\mu\nu}\nabla_AY_{\mu\nu}^A$. The second term matches with the one in four dimensions \cite{Liu:2023qtr} and we can rewrite it as 
\bea 
-\frac{1}{2}\omega^{\mu\nu}\int du d\Omega Y_{\mu\nu}^A (\dot{A}^B\nabla^CA^D-A^B\nabla^C\dot A^D)P_{ABCD},
\eea which is exactly the one  \eqref{MY} obtained from boundary Hamiltonian by taking normal order and flipping the sign.

In conclusion, the supertranslation and superrotation generators are actually the generalization of Poincar\'e fluxes from bulk to boundary. However, there is still an operator $\mathcal{O}_{\bm h}$ which is not well understood. In four dimensions, there is an additional electromagnetic duality \cite{Oliver:1892,Dirac:1931kp,Dirac:1948um,1957AnPhy...2..525M} in the bulk and the corresponding conserved current leads to the helicity flux. However, there is no analogous duality invariance in higher dimensions\footnote{In even dimensions, one can extend the electromagnetic duality by  antisymmetric tensor fields \cite{Nepomechie:1984wu,Teitelboim:1985ya}. However, this would introduce extended objects which is not the point particle discussed in this article. } and we should be much more careful  with this new operator.

\section{Microscopic interpretation}\label{int}
In this section, we will claim that the operator $\mathcal{O}_{\bm h}$ is actually the electromagnetic helicity flux operator using the method of canonical quantization, even though there is no corresponding duality invariance. We substitute \eqref{canonicalA} into the definition of $\mathcal{O}_{\bm h}$ 
\bea 
\mathcal{O}_{\bm h}&=&-i \int d^{d-1}\bm k\  h^{ab}b_{a,\bm k}^\dagger b_{b,\bm k},
\eea where 
\bea 
h^{ab}=h^{AB}e_{A}^{\ a}e_{B}^{\ b}
\eea whose inverse is 
\bea 
h_{AB}=h_{ab}e^a_A e^b_B.
\eea 
Given the metric \eqref{metricgamma}, we may choose the vielbeins as 
\begin{align}
    e^{a}_A=\begin{pmatrix}
        1&0&0&\cdots&0\\
        0&\sin\theta_1&0&\cdots&0\\
        0&0&\sin\theta_1\sin\theta_2&\cdots&0
        \\
      \vdots&\vdots&\vdots&\ddots&\vdots\\
        0&0&0&\cdots&\sin\theta_1\cdots\sin\theta_{m-1}
    \end{pmatrix}
\end{align} The antisymmetric matrix $h_{ab}$ can be regarded as a smooth 2-form field on the flat space $\mathbb{R}^{m}$. The  simplest 2-form field is the generator of the rotation group $SO(m)$ in this plane. Consider a rotation in the $i$-$j$ plane, the corresponding generator is \bea 
 \left(h_{(ij)}\right)_{ab}=\delta_{ai}\delta_{bj}-\delta_{aj}\delta_{bi},\quad a,b,i,j=1,2,\cdots,m.\label{hab}
\eea 
Now the operator $\mathcal{O}_{\bm h}$ is 
\bea 
\mathcal{O}^{(ij)}_{\bm h}=-i \int d^{d-1}\bm k \left(b^{\dagger}_{i,\bm k}b_{j,\bm k}-b^{\dagger}_{j,\bm k}b_{i,\bm k}\right)=-\int d^{d-1}\bm k \left(b_{\text{R},\bm k}^{(ij)\dagger}b_{\text{R},\bm k}^{(ij)}-b_{\text{L},\bm k}^{(ij)\dagger}b_{\text{L},\bm k}^{(ij)}\right),
\eea where 
\bea 
b_{\text{R},\bm k}^{(ij)}=\frac{1}{\sqrt{2}}(b_i+ib_j),\qquad b_{\text{L},\bm k}^{(ij)}=\frac{1}{\sqrt{2}}(b_i-ib_j)
\eea and $b_{\text{R}/\text{L},\bm k}^{(ij)\dagger}$ are their Hermite conjugates. This is the difference of the number of photons with left and right hand polarizations with respect to the $i$-$j$ plane. 
Compared with the result of four dimensions, we conclude that the operator $\mathcal{O}_{\bm h}$ is the helicity flux operator for these ${m(m-1)}/{2}$ independent choices of $\bm h$. We will call them  minimal helicity flux operators.

The helicity flux operators are  non-Abelian in higher dimensions.
They form a Lie algebra $so(m)$ since 
\bea 
[h_{ab},h_{cd}]=\delta_{ac}h_{bd}-\delta_{bc}h_{ad}-\delta_{ad}h_{bc}+\delta_{bd}h_{ac}.
\eea 
We notice that $SO(m)$ is also the little group for the irreducible massless representation of Poincar\'e group $ISO(1,d-1)$\footnote{More accurately, this is the short little group for helicity representation. Interested readers may find more details on this point in Appendix \ref{littlegroup}.}. 

Before we close this section, we emphasize that the field $\bm h$ can be any smooth  2-form fields on $S^{m}$ and therefore captures the angular distribution of the helicity flux density operator 
\bea 
O_{AB}(u,\Omega)=\frac{1}{2}:\dot{A}_AA_B-\dot{A}_BA_A:.
\eea In four dimensions, the quantity is proportional to the Levi-Civita tensor on $S^2$ and we may extract a parity odd helicity flux density operator 
\bea 
O(u,\Omega)=\epsilon^{AB}:\dot{A}_BA_A:.\label{gomega}
\eea  However, this is not possible in higher dimensions.
\section{Topological term}
In this section, we will derive the minimal helicity flux operators at $\mathcal{I}^+$ as a topological term. In four dimensions, this is also called the chiral memory effect \cite{Maleknejad:2023nyh}. At first, we will review the Chern-Simons term for electromagnetic theory. We will adopt the language of differential geometry to simplify notations in this section. Therefore, the vector field $a_\mu$ is denoted as a one-form 
\be 
a=a_\mu dx^\mu
\ee and the electromagnetic field $f_{\mu\nu}$ is a two-form field 
\be 
f=\frac{1}{2}f_{\mu\nu}dx^\mu\wedge dx^\nu=da.
\ee We may choose a hypersurface $\mathcal{H}$ in four dimensions and then the Chern-Simons term on this surface is 
\be 
I[a]=\int_{\mathcal{H}} a\wedge da.
\ee This term could be regarded as a boundary term from the second Chern character in the  bulk 
\bea 
S[a]=\int_{M} f\wedge f.
\eea Let us choose two different kinds  of surfaces  to discuss the Chern-Simons term. 
\begin{enumerate}
    \item The hypersurface $\mathcal{H}$ is the constant time slice. For simplicity, we can set $t=0$. Then 
    \bea 
    I[a]=\int_{\mathcal{H}} a_i dx^i \wedge \frac{1}{2}f_{jk}dx^j\wedge dx^k=\frac{1}{2}\int_{\mathcal{H}} a_i f_{jk}\epsilon^{ijk}d^3\bm x=\int_{\mathcal{H}} a_i b^i d^3\bm x,
    \eea where we have written the magnetic field $b^i$ as 
    \be 
    b^i=\frac{1}{2}\epsilon^{ijk}f_{jk}.
    \ee In terms of the language of three-dimensional vector analysis, we find the following magnetic helicity 
    \be 
    I[a]=\int_{\mathcal{H}}d^3\bm x \ \bm a\cdot\bm b.
    \ee It is understood that this  magnetic helicity measures the linkage of the magnetic field lines \cite{1958PNAS...44..489W}.
    \item In this paper, we are not interested in the hypersurface of constant time. Instead, we are considering the future/past null infinity ($\mathcal{I}^{\pm}$). Therefore, we will choose the hypersurface $\mathcal{H}=\mathcal{I}^+$. Technically, we may choose a constant $r$ slice $\mathcal{H}_r$ and then send $r\to\infty$ while keeping retarded time $u$ finite. In this case, the Chern-Simons term becomes 
    \bea 
    I[a]&=&\lim_{r\to\infty, u\ \text{finite}}\int_{\mathcal{H}_r} a_\mu dx^\mu\wedge \frac{1}{2}f_{\nu\rho}dx^\nu \wedge dx^\rho \nn\\&=&\frac{1}{2}\lim_{r\to\infty, u\ \text{finite}}\int_{\mathcal{H}_r}a_\mu f_{\nu\rho}\epsilon^{\mu\nu\rho\sigma}(d^3x)_\sigma\nn\\&=& -\frac{1}{2}\lim_{r\to\infty}\ r^2 \int du d\Omega a_\mu f_{\nu\rho}m_\sigma \epsilon^{\mu\nu\rho\sigma}\nn\\&=&{ -}\frac{1}{2}\int du d\Omega N_\mu^{\ \alpha} A_\alpha A_{\nu\rho}^{\ \ \beta}\dot{A}_{\beta} m_\sigma \epsilon^{\mu\nu\rho\sigma}.
    \eea

    From the definition of $N_\mu^{\ \alpha}$ in \eqref{Nd} and $A_{\nu\rho}^{\ \ \beta}$ in \eqref{abcd}, { and} the antisymmetric property of the Levi-Civita tensor, we obtain 
    \bea 
    I[a]&=&{ -}\frac{1}{2}\int du d\Omega A_\alpha \dot{A}_{\beta} \epsilon^{\mu\nu\rho\sigma}(-Y^A_\mu \delta^\alpha_A)(-Y^B_{\nu\rho}\delta^\beta_B)m_\sigma\nn\\&=&{ -}\int du d\Omega A_A \dot{A}_B \epsilon^{ij0k}Y_i^A Y^B_{j0}n_k\nn\\&=&-\int du d\Omega A_A\dot{A}_B \epsilon^{ijk}Y_i^A Y_j^B n_k\nn\\&=&-\int du d\Omega A_A \dot{A}_B \epsilon^{AB}.
    \eea This is exactly the helicity flux operator with $g=-1$ in four dimensions. 
\end{enumerate}
Now we will extend the discussion  to higher dimensions. A naive extension of the Chern-Simons term to higher dimensions is the $j$-th Chern character ($j>2$). However, it would contain higher derivative terms and doesn't match with the helicity flux operator constructed in this work. We may consider the following topological action in the bulk\footnote{ This kind of action has been introduced in \cite{Baulieu1997SpecialQF,Baulieu1997CohomologicalYT}.}
\bea 
S[a]=\int_M f\wedge f\wedge {\tt g}
\eea where $\tt g$ is a $d-4$ form
\bea 
{\tt g}=\frac{1}{(d-4)!}{\tt g}_{\mu_1\cdots \mu_{d-4}}dx^{\mu_1}\wedge \cdots\wedge dx^{\mu_{d-4}}.
\eea For this action to be a total derivative, we may choose $\tt g$ 
a constant $d-4$ form\footnote{Actually, the $d-4$ form only needs to be closed $d{\tt g}=0$.
}. Then the corresponding boundary term is \cite{Baulieu:1997nj}
\bea 
I[a]=\int_{\mathcal{H}} a\wedge da \wedge {\tt g}.
\eea Now we choose $\mathcal{H}=\mathcal{I}^+$ 
\bea 
I[a]&=&-\frac{(-1)^d}{2\times (d-4)!}\int du d\Omega A_\alpha \dot{A}_\beta N_\mu^{\ \alpha}A_{\nu\rho}^{\ \ \beta}{\tt g}_{\sigma_1\cdots\sigma_{d-4}}m_\sigma\epsilon^{\mu\nu\rho\sigma_1\cdots\sigma_{d-4}\sigma}\nn\\&=&-\frac{(-1)^d}{2\times (d-4)!}\int du d\Omega A_A \dot{A}_B Y_\mu^A Y_{\nu\rho}^B {\tt 
g}_{\sigma_1\cdots\sigma_{d-4}}m_\sigma \epsilon^{\mu\nu\rho\sigma_1\cdots\sigma_{d-4}\sigma}.
\eea In the first line, 
we have used the convention
    \begin{align}
  \epsilon^{\mu_1\cdots\mu_d}(d^{d-1}x)_{\mu_d}&=\epsilon^{\mu_1\cdots\mu_d}\times\frac{1}{(d-1)!}\epsilon_{\mu_d\nu_1\cdots\nu_{d-1}}dx^{\nu_1}\wedge \cdots\wedge dx^{\nu_{d-1}}\nn\\
  &=(-1)^ddx^{\mu_1}\wedge\cdots\wedge dx^{\mu_{d-1}}.
\end{align}
In the second line, we have used the antisymmetric  property of Levi-Civita tensor. Therefore, 
\bea 
I[a]&=&\frac{(-1)^d}{(d-4)!}\int du d\Omega A_A \dot{A}_B Y_i^A Y_j^B{\tt g}_{k_1\cdots k_{d-4}}m_l \epsilon^{ij0k_1\cdots k_{d-4}l}\nn\\&=&-\frac{(-1)^d}{(d-4)!}\int du d\Omega A_A \dot{A}_B Y_i^A Y_j^B n_l{\tt g}_{k_1\cdots k_{d-4}}\epsilon^{ijk_1\cdots k_{d-4}l}.
\eea Notice that the CKV  satisfies the completeness relation \eqref{compY}, we find the following completeness relation in the Euclidean space $\mathbb{R}^{d-1}$
\be 
Y_i^A Y_{jA}+n_i n_j=\delta_{ij}.
\ee We may choose the vielbein field 
\bea 
E_i^A=Y_i^A,\quad E_i^{\perp}=n_i
\eea where we have used $\perp$ to denote the direction of the normal vector of the celestial sphere $S^{d-2}$. Therefore, 
\bea 
\frac{1}{(d-4)!}Y_i^AY_j^B n_l {\tt g}_{k_1\cdots k_{d-4}}\epsilon^{ijk_1\cdots k_{d-4}l}&=&\frac{1}{(d-4)!} {
\tt g}_{k_1\cdots k_{d-4}}\epsilon^{ABk_1\cdots k_{d-4}\perp}=(-1)^{d-2}(*{\tt g})^{AB}.
\eea At the last step, we have  chosen spherical coordinates and then $k_i\not=\perp$. The Levi-Civita tensor in $\mathbb{R}^{d-1}$ reduces to the Levi-Civita tensor on the celestial sphere. The Hodge dual $*g$ is defined with respect to $d-2$ dimensional sphere 
\bea 
(*{\tt g})^{AB}=\frac{1}{(d-4)!}\epsilon^{ABC_1\cdots C_{d-4}}{\tt g}_{C_1\cdots C_{d-4}}.
\eea By identifying the two-form field $\bm h$ with the Hodge dual of $\tt g$, we find 
\bea 
I[a]=\int du d\Omega A^A\dot{A}^B h_{AB},
\eea which is exactly the helicity flux operator. The discussion can be extended to general null hypersurfaces. Interested reader can find the details in Appendix \ref{hyper}.

\section{Alternative derivations of the helicity flux operators} \label{alt}
In the previous sections, we find the electromagnetic helicity flux operators by calculating the commutators of the superrotation generators. In this section, we provide two alternative ways to obtain the same operators.
\subsection{As a boundary Hamiltonian}
In this subsection, the helicity flux operator is found through Hamilton's equation \eqref{hamiltonian} which needs a variation of the fundamental field. This variation may be read out from the commutator between the helicity flux operator and the fundamental field $A_A$  
\bea 
[\mathcal{O}_{\bm h},A_A(u,\Omega)]=i h_{AB}(\Omega)A^B(u,\Omega).\label{sd}
\eea This indicates that the associated covariant variation of $A_A$ is 
\bea 
\slashed\delta_{\bm h} A_A=\Delta_A(\bm h;A;u,\Omega)=-h_{AB}(\Omega)A^B(u,\Omega)\label{superduality}
\eea which is the higher dimensional analog of the superduality transformation in four dimensions. 
Therefore we can  obtain the corresponding Hamiltonian using Hamilton's equation \eqref{hamiltonian}
\bea 
H_{\bm h}=\int du d\Omega h_{AB}\dot{A}^BA^A,
\eea which matches with the form of the helicity flux operator after quantization.

\subsection{As an extension of duality rotation generators} 
Though there is no exact electromagnetic duality invariance in the bulk, we may still provide a rather formal bulk duality transformation with which one can derive the helicity flux. 
Following the logic in the 4-dimensional case, we construct a duality-symmetric action
\begin{align}
  S=-\frac{1}{8}\int d^dx[f_\mn f^\mn+\widetilde f_\mn \widetilde f^\mn],\label{dualact}
\end{align}
where we have introduced another 1-form $\widetilde a$ and it field strength tensor
\begin{align}
  \widetilde f=d\widetilde a \quad\Rightarrow\quad  \widetilde f_\mn=\p_\m \widetilde a_\nu-\p_\n \widetilde a_\mu.
\end{align}
These two theories are not coupled to each other, and it is easy to find the equations of motion
\begin{align}
    df=d*f=0,\qquad d\widetilde f=d*\widetilde f=0.
\end{align}
The action and equations of motion are invariant under the following duality rotation
\begin{align}
  \begin{split}
      f_\mn'&= f_\mn\cos\vp+ \widetilde f_\mn\sin\vp,\\ \widetilde f_\mn'&= -f_\mn\sin\vp+ \widetilde f_\mn\cos\vp.
  \end{split}\label{fmn'}
\end{align}
The $SO(2)$ parameter $\vp$ is a constant and thus \eqref{fmn'} gives
\begin{align}
  a_\m'= a_\m\cos\vp+ \widetilde a_\m\sin\vp,\qquad  \widetilde a_\m'= -a_\m\sin\vp+ \widetilde a_\m\cos\vp,
\end{align}
from which the infinitesimal duality transformation is
\begin{align}
  \delta_{\epsilon}a_\mu=\epsilon\widetilde a_\mu,\qquad \delta_{\epsilon}\widetilde a_\mu=-\epsilon a_\mu.\label{bulkdual}
\end{align}
Until now, we do not know what is the field $\widetilde a$ and whether (or how) it is related to the original field $a$. As long as we construct \eqref{dualact}, the above symmetry holds. What we actually do is to generalize the concept of duality between two well-known or related theories (e.g. electromagnetic duality in 4 dimensions) to duality between a theory and its similar but not directly relevant counterpart.

The conserved current for \eqref{bulkdual} is easy to find
\begin{align}
  j_{\rm duality}^\mu&=-\frac{\partial \cl}{\partial (\partial_\mu a_\nu)}\delta a_\nu-\frac{\partial \cl}{\partial (\partial_\mu \widetilde a_\nu)}\delta \widetilde a_\nu\nn\\
  &=\frac{1}{2}(f^{\mu\nu} \delta a_\nu+\widetilde f^{\mu\nu} \delta \widetilde a_\nu)=\frac{1}{2}(f^{\mu\nu}\widetilde{a}_\nu-\widetilde{f}^{\mu\nu}a_\nu).
\end{align}
Imposing the same fall-offs for $\widetilde a_\mu$ as \eqref{falloffamu0}, we find
\begin{align}
  j_{\rm duality}^{\mu}&=\frac{1}{2}r^{-2\Delta}[(-\frac{1}{2}n^\mn \dot A_r-Y^{\mn A} \dot A_A)N^\a_\n\widetilde{A}_\a-(-\frac{1}{2}n^\mn \dot{\widetilde A}_r-Y^{\mn A} \dot{\widetilde A}_A)N^\a_\n{A}_\a]+\cdots.
\end{align}
With the help of the identities
\begin{align}
  n^\mn N^\a_\n&=(n^\mu\bar{n}^\nu-n^\nu\bar{n}^\mu)(-n_\nu\delta^\alpha_u+m_\nu\delta^\alpha_r-Y_{\nu}^{A}\delta^\alpha_A)=-2\delta^\alpha_u n^\m+2\delta^\alpha_r\bar m^\m,\\
  Y_A^\mn N^\a_\n&=(Y_A^\mu {n}^\nu-Y^\nu_A{n}^\mu)(-n_\nu\delta^\alpha_u+m_\nu\delta^\alpha_r-Y_{\nu}^{B}\delta^\alpha_B)=Y^\m_A\delta^\alpha_r+n^\m \delta^\a_A,
\end{align}
one can compute
\begin{align}
  j_{\rm duality}^{\mu}&=\frac{1}{2}r^{-2\Delta}[ \dot A_r(m^\m\widetilde{A}_u-\bar m^\m\widetilde{A}_r)- \dot A_A(Y^{\m A}\widetilde{A}_r+n^\m \widetilde{A}^A)\nn\\
  &\quad -\dot{\widetilde A}_r(m^\m{A}_u-\bar m^\m{A}_r)+ \dot{\widetilde A}_A(Y^{\m A}{A}_r+n^\m {A}^A)]+\cdots,
\end{align}
Recalling the equality \(\dot A_r=0\) and the same for the dual field, we can derive the following flux 
\begin{align}
  \mathcal{F}_{\rm duality}&=r^{d-2}\int dud\Omega m_\m j_{\rm duality}^\mu\nn\\
  &=\frac{1}{2}\int dud\Omega m_\m n^\m(A_A\dot {\widetilde{A}}{}^A-\dot A_A\widetilde{A}^A)\nn\\
  &=-\int dud\Omega \dot A_A{\widetilde{A}}^A,\label{AatildeAa}
\end{align}
which  is a common result for duality symmetry just as in 4 dimensions and is manifest duality invariant. 

In the 4-dimensional duality-symmetric theory, the field $a_\mu$ has the same footing as the dual field $\widetilde a_\mu$. 
However, in higher dimensions, electric field $f_{0i}$ does not have the same number of degrees of freedom as the magnetic field $\frac{1}{2}\epsilon_{0k_3\cdots k_{d}ij}f_{ij}$, and thus there is no such a duality. But we can still write such a Lagrangian without the Hodge duality relation $\widetilde f_\mn=-*f_\mn$ as a constraint from which one can derive the boundary Hodge duality $\widetilde A_A=*A_A$ in 4 dimensions. Instead, we may construct the boundary ``dual" field \eqref{eq5.4}, and then demand a bulk field that can be used to formally construct a duality-symmetric action.

Now $\widetilde f_\mn$ is not the bulk Hodge dual of $f_\mn$ and $\widetilde A_A$ is not the boundary Hodge dual of $A_A$ in general dimensions. However, recalling what we did in section \ref{int}, we may introduce $h_{AB}$ as a rotation matrix associated with a transverse plane labeled by an antisymmetric constant tensor $\varpi^{ij}$
\be 
h_{AB}=\frac{1}{2}\varpi^{ij}\left(h_{(ij)}\right)_{ab}e^a_Ae^b_B=\frac{1}{2}\varpi^{ij}\left(h_{(ij)}\right)_{AB}.
\ee 
Apparently, we have $m(m-1)/2$ independent $\left(h_{(ij)}\right)_{AB}$ and thus the same number of dual fields which form the vector representation of $SO(m)$. It is obvious that in 4 dimensions, $\varpi^{ij}\propto \epsilon^{ij}$ which leads to the  well-defined boundary dual field \cite{Liu:2023qtr}
\be 
\widetilde{A}_A=-\epsilon_{AB}A^B.
\ee From the view of little group, since $SO(2)$ generalize to $SO(m)$, a unique boundary dual field should extend to $m(m-1)/2$ independent ones\footnote{Here, we assume the equivalence between the little group and boundary duality rotation for the massless vector. They are both $SO(m)$. When described on the celestial sphere $S^m$, both of them have a special direction such that the rotation is $SO(m)$ but not $SO(m+1)$. For little group, it is the direction of momentum, and for boundary duality rotation, it is the radial direction since we get it from large $r$ expansion of bulk duality. For massless particles, we often identify these two directions, at least in the context of scattering amplitudes. 

}.
A rotation of the field $A_A$ may be regarded as a dual field, thus we may impose a further condition 
\begin{align}
  \widetilde A_A=-h_{AB}A^B,\label{eq5.4}
\end{align}
and  get the helicity flux that we want
\begin{align}
    \mathcal{F}_{\rm duality}=\int dud\Omega A_Ah^{AB}\dot A_B.
\end{align}

As a consistency check, one can use the helicity flux operator to generate the boundary duality transformation 
\begin{align}
    [\mathcal{O}_{\epsilon\bm h},A_A(u,\Omega)]=i\epsilon h_{AB}A^B(u,\Omega)=-i\epsilon \widetilde A_A(u,\Omega)=-i\delta_\epsilon A_A(u,\Omega),
\end{align}
which agrees with \eqref{bulkdual}.

The above argument is a bit ad hoc, and there is some gap that we can not recover the complete bulk field from the boundary fundamental field alone. More explicitly, one can not get a unique $\widetilde a_\mu$ since $Y^\mu_A$ is not an invertible matrix. At last, we find something good in the above higher-dimensional derivation. It is about the angle-dependent generalization of the duality transformation. In higher dimensions, the angle dependence of $h_{AB}(\Omega)$ comes from the construction of the dual field in \eqref{eq5.4}, but has nothing to do with the generalization of $SO(2)$ rotation \eqref{fmn'} which does not hold as a symmetry for the bulk theory. In 4 dimensions, there is a natural dual field $\widetilde A_A=-\ep_{AB}A^B$ which is consistent with the bulk reduction and Hodge duality, and so we have no reason to add another angle-dependent factor for it, i.e., write $\widetilde A_A=-g(\Omega)\ep_{AB}A^B$.


\section{Conclusion and discussion}\label{conc}
In this work, we have constructed the electromagnetic helicity flux operator in higher dimensions. This operator is added to the energy flux and angular  momentum flux operator to form a closed algebra. The algebra extends the one in four dimensions due to the non-Abelian feature of the helicity flux operator. We have checked the interpretation of the helicity flux operator by transforming it into the momentum  space using mode expansion. We have also used a Chern-Simons like action on the Carrollian manifold to find the same helicity flux operator. 
Despite the non-Abelian property of the helicity flux operator, there are more interesting topics that deserve further study.

Firstly, the gravitational helicity flux operator in higher dimensions could be found after some effort. One of the key observations  in this work is that the transformation law of the fundamental field is formally the same in general dimensions. This has been checked for scalar theory \cite{Li:2023xrr} and electromagnetic theory in this work. It seems to be true even for gravitational theory as it reflects the intrinsic property of the Carrollian manifold. The gravitational helicity flux operator
\begin{align}
    \mathcal{O}_{\bm h}^{(s=2)}=\int dud\Omega \dot C_{AC} C^C_{\ B} h^{BA}(\Omega)
\end{align}
can be read out from the commutators straightforwardly without solving the complicated Einstein equation.  It would be nice to work out the details and check this point in the future.

Secondly, there is no known electromagnetic duality invariance associated with the helicity flux operator in the bulk. At the null boundary, the helicity flux operator indeed generates a  similar superduality transformation \eqref{sd} that preserves the symplectic form.  The relation between the Carrollian diffeomorphism and the energy and angular momentum fluxes has been argued in \cite{Liu:2024nkc,Liu:2024nfc} where one of the key ingredients is the conservation of the stress tensor. The arguments cannot be applied directly to the  higher dimensional helicity flux operator since we lack corresponding conserved currents. Interestingly, in a paper that will be presented soon, we show that one can combine the vielbein field $e_a^A$ and the fundamental field $A_A$ to form a radiative field $A_a=e_a^A A_A$ in the local Cartesian frame. 
One can rotate the vielbein field $e_a^A$, and consequently the field $A_a$ in the local frame by any element of the little group. This local rotation $SO(d-2)$ is exactly isomorphic to the action by the helicity flux operator defined in this paper. 

Thirdly, we have shown that the energy flux, angular momentum flux and the helicity flux operators $\{\mathcal{T}_f,\mathcal{M}_{\bm Y},\mathcal{O}_{\bm h}\}$ form a closed algebra. We notice that the angular momentum flux operator $\mathcal{M}_{\bm Y}$ can be deformed to a new operator
\be 
\widetilde{\mathcal{M}}_{\bm Y}=\mathcal{M}_{\bm Y}+\mathcal{O}_{\bm\tau}
\ee and then $\{\mathcal{T}_f,\widetilde{\mathcal{M}}_{\bm Y},\mathcal{O}_{\bm h}\}$ form an equivalent set of generators. The unspecified 2-form field $\bm\tau$ remains arbitrary and it may be interpreted as the torsion term in the definition of the covariant connection. When the torsion vanishes, the $\widetilde{\mathcal{M}}_{\bm Y}$ is the standard angular momentum flux operator. However, another choice of the torsion 
\be 
\bm\tau=\frac{1}{2}d\bm Y\label{torsiondY}
\ee is also promising since it subtracts exactly the terms which looks like the helicity flux operator in $\mathcal{M}_{\bm Y}$ and then 
\be 
\widetilde{\mathcal{M}}_{\bm Y}=\int du d\Omega \dot{A}^A(Y^C\nabla_CA_A+\frac{1}{2}\nabla_CY^C A_A)\label{deformedMY}
\ee is exactly the same form of the superrotation generator of the scalar theory, except that the scalar field $\Sigma$ is replaced by the vector field $A_A$. 
The structure still holds for general massless theories with nonvanishing spin. For example, with the torsion \eqref{torsiondY}, the superrotation generator $\mathcal{M}_{\bm Y}$ in the gravitational theory is deformed to 
\bea 
\widetilde{\mathcal{M}}_{\bm Y}=\int du d\Omega\  \dot{C}^{AB}(Y^C\nabla_CC_{AB}+\frac{1}{2}\nabla_CY^C C_{AB}),
\eea whose form is exactly the same as \eqref{deformedMY}.
Usually, the angular momentum of the particles is the summation of the orbital angular momentum and the ``intrinsic'' spin. In \cite{Palmerduca:2024goi}, the authors proved that there are both geometric and topological obstructions to prevent the decomposition of angular momentum for massless bosons to orbit and spin parts which form  representations of $SO(3)$ in four dimensions. This is not contradictory with our decomposition at the boundary since the minimal helicity flux operators form a representation of the little group but not of the spatial rotation. It would be better to understand whether this modified operator \eqref{deformedMY} could be interpreted as the orbital angular momentum in any sense. 

Finally, we only discussed the fluxes that cross $\mathcal I^+$ in this work. From a Carrollian field theory point of view, one can insert local operators, e.g., flux densities, at the null boundary. Therefore, the flux operator is just a smeared operator that contains equivalent information on the flux densities. The fluxes are finite in the radiative fall-off condition. However, the bare symplectic potential and surface charge are divergent in general. Fortunately, one may use the ambiguities of the symplectic potential to regularize the charges. The procedure is partly similar to the holographic renormalization \cite{deHaro:2000vlm,Skenderis:2002wp}. In general relativity, it has been successfully applied to renormalize the superrotation charge \cite{Compere:2018ylh}. Similar work can be found in \cite{Freidel:2019ohg} for higher dimensional electromagnetic theory. It is believed that the time derivative of the surface charge obeys a flux formula \cite{Wald:1999wa,Compere:2020lrt}. Given the fluxes presented in this work, one could expect a finite surface charge after regularizing it carefully.  

\paragraph{Notes added.}
Recently, we found that the ``central charge'' ${\rm C}_T$ will lead to a violation of Jacobi identity \cite{Guo:2024qzv}
\begin{align}
  J[\ct_{f_1},\ct_{f_2},\cm_{\bm Y}]
  &=-\frac{m}{48\pi}\delta^{(m)}(0)\ci_{Y^A\nabla_A(f_1\p_u^3f_2-f_2\p_u^3f_1)}.
\end{align}
This violation implies that the fluxes corresponding to the general supertranslation $f(u,\Omega)$ and superrotation $Y^A(\Om)$ generally can not form a Lie algebra through the commutators.

To cure this problem, we can either demand $f$ to be at most quadratic about $u$ or constrain $\bm Y$ to be a (smooth) divergence-free vector field (DVF) on the sphere. For the first solution, ${\rm C}_T$ vanishes and we can further constrain $f$ to get the (higher-dimensional version of) Weyl BMS \cite{Freidel:2021fxf} and generalized BMS algebras \cite{Campiglia:2014yka,Campiglia:2015yka} both of which will still be intertwined with the helicity flux.
For the second solution, the Jacobi identity violation becomes a total derivative on the sphere and thus vanishes. We should check that DVFs on the sphere are closed under commutators
\begin{align}
\nabla_A[\bm Y,\bm Z]^A=  Y^A\nabla_A\nabla_B Z^B-Z^B\nabla_B\nabla_AY^A=0,\label{ynnza.34}
\end{align}
where only the properties of the Riemann tensor are used so it is a general result. More specifically, we need to use
\begin{align}
    [\nabla_A,\nabla_B] V^B=-R_{AB}V^B 
\end{align}
and $R_{AB}=R_{BA}$. In this case, the central charge ${\rm C}_T$ can survive the Jacobi identity. We call what smooth DVFs on the 2-sphere generate magnetic superrotation \cite{Guo:2024qzv} since it has the magnetic (or odd) parity \cite{Flanagan:2015pxa} in four dimensions, i.e., $Y^A=\epsilon^{AB}\nabla_B\mathcal{Y}$ (such as spatial rotation generator $Y_{ij}^A$, seeing appendix \ref{identities} for its definition). In four dimensions, any vector field on the 2-sphere can always be decomposed into a DVF and a rotational-free part \cite{Compere:2018ylh} with electric (or even) parity (e.g., Lorentz boost generator $Y_{0i}^A$). In general dimensions, a DVF could be determined by $m-1$ scalar functions, consistent with the situation in four dimensions. 
Both ways to cure the Jacobi identity violation do not affect the appearance of helicity flux operator. Therefore, the main concern of this paper, e.g., the construction and interpretation of helicity flux operator, are still valid.

\vspace{10pt}
{\noindent \bf Acknowledgments.} 
The work of J.L. was supported by NSFC Grant No. 12005069.  The work of W.-B. Liu and X.-H. Zhou is supported by ``the Fundamental Research Funds for the Central Universities'' with No. YCJJ20242112.

\appendix
\section{Identities}\label{identities}
There are various identities that  are useful in the derivation. We will collect these  identities in this appendix though some of the identities have already appeared in \cite{Li:2023xrr,Liu:2023jnc}. The unit normal vector of the $m=d-2$ dimensional sphere is denoted as $n^i,\quad i=1,2,\cdots,d-1$
\be 
n^i n^i=1.
\ee The null vectors $n^\mu$ and $\bar n^\mu$ are 
\bea 
n^\mu=(1,n^i),\quad \bar n^\mu=(-1,n^i). 
\eea 
We may construct three quantities through $n^\mu$ and $\bar n^\mu$
\bea 
m^\mu=\frac{1}{2}(n^\mu+\bar n^\mu)=(0,n^i),\quad \bar m^\mu=\frac{1}{2}(n^\mu-\bar n^\mu)=(1,0),\quad Y^\mu_A=-\nabla_A n^\mu.
\eea The last quantity $Y^\mu_A$ could be regarded as a $d$-vector in Minkowski spacetime or a $m=d-2$ dimensional vector on $S^m$. One can use $\eta_{\mu\nu}$ to lower its Greek index to obtain 
\bea 
Y_{\mu A}=\eta_{\mu\nu}Y^\nu_A.
\eea Similarly, one can also use $\gamma^{AB}$ to raise its Latin index 
\bea 
Y^{\mu A}=\gamma^{AB}Y^\mu_B.
\eea 
The $d$-vectors $n^\mu,\bar n^\mu, Y^\mu_A$ form a complete basis to expand any vector field. They satisfy the orthogonality relations 
\bea 
&&n^2=0=\bar n^2,\quad n\cdot \bar n=2,\quad Y^\mu_A Y_{\mu B}=\gamma_{AB},\quad n^\mu Y_\mu^A=\bar n^\mu Y_\mu^A=0
\eea and the completeness relation 
\bea 
Y_\mu^A Y_{\nu A}+\frac{1}{2}(n_\mu \bar n_\nu+n_\nu\bar n_\mu)=\eta_{\mu\nu}.\label{compY}
\eea 
With these identities, we can find the orthogonality relations involving $m^\mu$ and $\bar m^\mu$
\bea 
&& m^2=1,\quad \bar m^2=-1,\quad m\cdot \bar m=0,\quad m^\mu Y^A_\mu=\bar m^\mu Y^A_\mu=0.
\eea The completeness relation becomes 
\bea 
Y_\mu^A Y_{\nu A}+m_\mu m_\nu-\bar m_\mu\bar m_\nu=\eta_{\mu\nu}.
\eea One may check that $Y_\mu^A,\quad \mu=1,2,\cdots,d-1$ are $m+1$ strictly conformal Killing vectors (CKVs) on $S^m$
\bea 
\nabla_A Y^\mu_B+\nabla_BY^\mu_A=\frac{2}{m}\gamma_{AB}\nabla_CY^{\mu C}.
\eea By definition, we also find 
\bea 
\nabla_AY^\mu_B=\nabla_BY^\mu_A=\frac{1}{m}\gamma_{AB}\nabla_CY^{\mu C}=\gamma_{AB}m^\mu,
\eea 
where we have used the identity 
\bea 
\nabla^A\nabla_A n_\mu=-\nabla^AY_{\mu A}=-m\ m_\mu.
\eea 
The ${(m+2)(m+1)}/{2}$ CKVs on $S^m$ are 
\bea 
Y^{\mu\nu}_A=Y^\mu_A n^\nu-Y^\nu_A n^\mu
\eea in which 
$
Y^{0i}_A=-Y^i_A
$ are strictly CKVs and $Y^{ij}_A=Y^i_A n^j-Y^j_A n^i$ are Killing vectors (KVs) 
\bea 
\nabla_A Y^{ij}_B+\nabla_B Y^{ij}_A=0.
\eea 
Note that one may also construct the following CKVs 
\bea 
\bar{Y}^{\mu\nu}_A=Y^\mu \bar{n}^\nu-Y^\nu_A\bar n^\mu
\eea which relate to $Y^{\mu\nu}_A$ through 
\bea 
\bar{Y}^{0i}_A=Y^i_A=-Y^{0i}_A,\quad \bar{Y}^{ij}_A=Y^{ij}_A.
\eea 
Some identities related to $Y_\mu^A$ or $Y_{\mu\nu}^A$ are
\bea 
Y_\mu^A Y^{\mu\nu}_A=m n^\nu,\quad Y_\mu^A \bar{Y}^{\mu\nu}_A=m\bar n^\nu,\quad Y_{\mu\nu}^AY_{\rho\sigma A}={ \gamma_{\mu\rho}n_\nu n_\sigma-\gamma
_{\mu \sigma }n_\nu n_\rho-\gamma_{\nu\rho}n_\mu n_\sigma+\gamma_{\nu \sigma }n_\mu n_\rho}.
\eea 
It is not hard to prove that 
\bea 
n_{\mu\nu}\equiv n_\mu \bar n_\nu-n_\nu \bar n_\mu=2(n_\mu m_\nu-n_\nu m_\mu)=-\frac{2}{m}\nabla_AY^A_{\mu\nu}\quad\Rightarrow\quad \nabla_AY^A_{\mu\nu}=-\frac{m}{2}n_{\mu\nu},
\eea 
and therefore
\bea 
\nabla_A n^{\mu\nu}=Y^{\mu\nu}_A-\bar{Y}^{\mu\nu}_A.
\eea 
We define the following tensors
\bea 
m^A_{\mu\nu}&=&Y_\mu^A m_\nu-Y_\nu^A m_\mu=\frac{1}{2}(Y_{\mu\nu}^A+\bar Y_{\mu\nu}^A),\\
Y_{\mu\nu}^{AB}&=&Y_\mu^AY_\nu^B-Y_\nu^AY_\mu^B.
\eea 
There are various identities that are useful 
\bs\begin{align}
& n^\mu n_{\mu\nu}=-2n_\nu,\quad n^\mu Y_{\mu\nu}^A=0,\quad n^\mu m_{\mu\nu}^A=-Y_\nu^A,\quad n^\mu Y_{\mu\nu}^{AB}=0,\\
& m^\mu n_{\mu\nu}=-2\bar m_\nu,\quad m^\mu Y_{\mu\nu}^A=-Y_\nu^A,\quad m^\mu m_{\mu\nu}^A=-Y^A_\nu,\quad m^\mu Y_{\mu\nu}^{AB}=0,\\
& Y^{\mu A} n_{\mu\nu}=0,\quad Y^{\mu A}Y_{\mu\nu}^B=\gamma^{AB}n_\nu,\quad Y^{\mu A} m_{\mu\nu}^B=\gamma^{AB}m_\nu,\\ & Y^{\mu C}Y_{\mu\nu}^{AB}=\gamma^{CA}Y^B_\nu-\gamma^{CB}Y^A_\nu,\\
& n_{\mu\rho}n_\nu^{\ \rho}=-2(n_\mu\bar n_\nu+\bar n_\mu n_\nu),\quad n_{\mu\rho}Y_\nu^{\ \rho A}=2n_\mu Y^A_\nu,\\&n_{\mu\rho}m_{\nu}^{\ \rho A}=2\bar m_\mu Y_\nu^A,\quad n_{\mu\rho}Y_{\nu}^{\ \rho AB}=0,\\& Y_{\mu\rho}^A m_\nu^{\ \rho B}=Y^A_\mu Y^B_\nu+\gamma^{AB}n_\mu m_\nu, \quad Y_{\mu\rho}^A Y_\nu^{\ \rho B}=\gamma^{AB}n_\mu n_\nu,\\& Y^A_{\mu\rho}Y_{\nu}^{\ \rho BC}=-\gamma^{AC}Y_\nu^Bn_\mu+\gamma^{AB}Y_\nu^C n_\mu,\\
& Y_{\mu\rho}^{AB}m_{\nu}^{\ \rho C}=-\gamma^{BC}Y^A_\mu m_\nu+\gamma^{AC}Y_\mu^B m_\nu,\\ &Y_{\mu\rho}^{AB}Y_{\nu}^{\ \rho CD}=\gamma^{BD}Y_\mu^A Y^C_\nu-\gamma^{BC}Y_\mu^A Y_\nu^D-\gamma^{AD}Y_\mu^B Y_\nu^C+\gamma^{AC}Y_\mu^B Y_\nu^D,\\ & m_{\mu\rho}^Am_{\nu}^{\ \rho B}=Y_\mu^AY_\nu^B+\gamma^{AB}m_\mu m_\nu.
\end{align}\es
 We have also defined the tensors $N_\mu^{\ \alpha}$ and $\bar N_\alpha^{\ \mu}$ which obey the following identities
 \bea 
&&m^\mu N_\mu^{\ \alpha}=-\delta^\alpha_u+\delta^\alpha_r,\quad n^\mu N_\mu^{\ \alpha}=\delta^\alpha_r,\quad 
 Y_\mu^A N^{\mu\alpha}=-\delta^\alpha_A,\\
 &&N_\nu^{\ \alpha}N^{\nu\beta}=-\delta^\alpha_u\delta^\beta_r-\delta^\alpha_r\delta^\beta_u+\delta^\alpha_r\delta^\beta_r+\gamma^{AB}\delta^\alpha_A\delta^\beta_B.
\eea 

\paragraph{Identities for equation of motion.} We may define the following four tensors 
\bs\label{abcd}\begin{align}
A_{\mu\nu}^{\ \ \alpha} & \equiv n_\nu N_\mu^{\ \alpha}-n_\mu N_{\nu}^{\ \alpha}=-\frac{1}{2}n_{\mu\nu}\delta^\alpha_r-Y^A_{\mu\nu}\delta^\alpha_A,\\
 B_{\mu\nu}^{\ \ \alpha} &\equiv m_\nu N_\mu^{\ \alpha}-m_\mu N_\nu^{\ \alpha}=-\frac{1}{2}n_{\mu\nu}\delta^\alpha_u-m^A_{\mu\nu}\delta^\alpha_A,\\
C_{\mu\nu}^{\ \ \alpha} &\equiv -Y_\mu^A \nabla_A N_\nu^{\ \alpha}+Y_\nu^{\ A}\nabla_A N_\mu^{\ \alpha}=m_{\mu\nu}^A\delta^\alpha_A,\\
D_{\mu\nu}^{\ \ \alpha A} &\equiv -Y_\mu^A N_\nu^{\ \alpha}+Y^A_\nu N_\mu^{\ \alpha}=Y^A_{\mu\nu}\delta^\alpha_u- m^A_{\mu\nu}\delta^\alpha_r+Y_{\mu\nu}^{AB}\delta^\alpha_B.
 \end{align}
 \es
 Then, we have
 \bs\begin{align}
 n^\mu A_{\mu\nu}^{\ \ \alpha}&=n_\nu\delta^\alpha_r,\\
 n^\mu B_{\mu\nu}^{\ \ \alpha}&=n_\nu\delta^\alpha_u+Y^A_\nu\delta^\alpha_A,\\
 n^\mu C_{\mu\nu}^{\ \ \alpha}&=-Y_\nu^A\delta^\alpha_A,\\
 n^\mu D_{\mu\nu}^{\ \ \alpha A}&=Y^A_\nu \delta^\alpha_r,
 \end{align}
 \es
 \bs\begin{align}
    m^\mu A_{\mu\nu}^{\ \ \alpha}&=\bar m_\nu\delta^\alpha_r+Y_\nu^A\delta^\alpha_A,\\
 m^\mu B_{\mu\nu}^{\ \ \alpha}&=\bar m_\nu\delta^\alpha_u+Y^A_\nu\delta^\alpha_A,\\
 m^\mu C_{\mu\nu}^{\ \ \alpha}&=-Y_\nu^A\delta^\alpha_A,\\
 m^\mu D_{\mu\nu}^{\ \ \alpha A}&=-Y^A_\nu (\delta^\alpha_u-\delta^\alpha_r),
 \end{align}
 \es 
 \bs\begin{align}
    Y_A^\mu A_{\mu\nu}^{\ \ \alpha}&=-n_\nu \delta^\alpha_A,\\
 Y_A^\mu B_{\mu\nu}^{\ \ \alpha}&=-m_\nu\delta^\alpha_A,\\
 Y_A^\mu C_{\mu\nu}^{\ \ \alpha}&=m_\nu\delta^\alpha_A,\\
 Y_A^\mu D_{\mu\nu}^{\ \ \alpha B}&=\delta^B_A(n_\nu\delta^\alpha_u-m_\nu\delta^\alpha_r)+(\gamma^{B}_AY_\nu^C-\gamma^{C}_AY_\nu^B)\delta^\alpha_C,
 \end{align}
 \es 
 \bs\begin{align}
 &Y^\mu_AY^\nu_B A_{\mu\nu}^{\ \alpha}=Y^\mu_AY^\nu_B B_{\mu\nu}^{\ \alpha}=Y^\mu_AY^\nu_B C_{\mu\nu}^{\ \alpha}=0,\quad Y^\mu_AY^\nu_B D_{\mu\nu}^{\ \alpha C}=(\gamma_A^C\gamma_B^D-\gamma_A^D\gamma_B^C)\delta^\alpha_D,\\
 & Y^\nu_Bm^\mu A_{\mu\nu}^{\ \alpha}=Y^\nu_Bm^\mu B_{\mu\nu}^{\ \alpha}=-Y^\nu_Bm^\mu C_{\mu\nu}^{\ \alpha}=\delta^\alpha_B,\quad Y^\nu_Bm^\mu D_{\mu\nu}^{\ \alpha A}=-\delta^A_B(\delta^\alpha_u-\delta^\alpha_r),\\ &\bar m^\mu n^\nu A_{\mu\nu}^{\ \ \alpha}=\delta^\alpha_r,\quad \bar m^\mu n^\nu B_{\mu\nu}^{\ \ \alpha}=\delta^\alpha_u,\quad \bar m^\mu n^\nu C_{\mu\nu}^{\ \ \alpha}=\bar m^\mu n^\nu D_{\mu\nu}^{\ \ \alpha A}=0.
 \end{align}\es  As a consequence, we find 
 \bs\begin{align}
     n_\mu \dot{f}^{\mu\nu (k)}=&\ n^\nu \ddot{A}^{(k)}_r+(\Delta+k-1)n^\nu \dot{A}_u^{(k-1)}+(\Delta+k-2)Y^{\nu A}\dot{A}_A^{(k-1)}+Y^{\nu A}\nabla_A \dot{A}_r^{(k-1)},\\
     m_\mu f^{\mu\nu (k-1)}=&\ \bar m^\nu\dot{A}_r^{(k-1)}+Y^{\nu A}\dot{A}_A^{(k-1)}+(\Delta+k-2)\bar m^\nu A_u^{(k-2)}+(\Delta+k-3)Y^{\nu B}A^{(k-2)}_B\nn\\&-Y^{\nu B}\nabla_B A_u^{(k-2)}+Y^{\nu B}\nabla_BA_r^{(k-2)},\\
     Y_{\mu A} f^{\mu\nu (k-1)}=&-n^\nu\dot{A}^{(k-1)}_A-(\Delta+k-3)m^\nu A_A^{(k-2)}+n^\nu \nabla_A A_u^{(k-2)}-m^\nu \nabla_AA_r^{(k-2)}\nn\\&+Y^{\nu C}(\nabla_AA_C^{(k-2)}-\nabla_CA_A^{(k-2)}),\\
     \nabla^A (Y_{\mu A}f^{\mu\nu(k-1)})=&\ Y^{\nu A}\dot{A}_A^{(k-1)}-n^\nu\nabla^A\dot{A}_A^{(k-1)}+(\Delta+k-3)Y^{\nu A}A_A^{(k-2)}-(\Delta+k-3)m^\nu \nabla^AA_A^{(k-2)}\nn\\&-Y^{\nu A}\nabla_AA_u^{(k-2)}+n^\nu\nabla^2 A_u^{(k-2)}+Y^{\nu A}\nabla_A A_r^{(k-2)}-m^\nu \nabla^2 A_r^{(k-2)}\nn\\&+Y^{\nu C}\nabla^A(\nabla_AA_C^{(k-2)}-\nabla_CA_A^{(k-2)}).
 \end{align}
 \es 
 We may also find 
 \bs\begin{align}
     Y_\nu^Am_\mu f^{\mu\nu(k-1)}&=\dot{A}^{A(k-1)}+(\Delta+k-3)A^{A(k-2)}-\nabla^AA_u^{(k-2)}+\nabla^A A_r^{(k-2)},\\
     Y_{\nu B}Y_{\mu A}f^{\mu\nu(k-1)}&=\nabla_AA_B^{(k-2)}-\nabla_BA_A^{(k-2)}.
 \end{align}\es 
 
 \paragraph{Identities for stress tensor.}
 Similarly, we can also work out the following  quadratic products
 \bs\begin{align}
     A_{\mu\rho}^{\ \ \alpha}A_{\nu}^{\ \rho\beta}=&\ -\frac{1}{2}(n_\mu\bar n_\nu+n_\nu\bar n_\mu)\delta^\alpha_r\delta^\beta_r+n_\mu Y_\nu^B \delta^\alpha_r \delta^\beta_B+n_\nu Y_\mu^A \delta^\beta_r\delta^\alpha_A+\gamma^{AB}n_\mu n_\nu \delta^\alpha_A\delta^\beta_B,\\
     A_{\mu\rho}^{\ \ \alpha}B_{\nu}^{\ \rho\beta}=&\ -\frac{1}{2}(n_\mu\bar n_\nu+n_\nu\bar n_\mu)\delta^\alpha_r \delta^\beta_u+\delta^\alpha_r\delta^\beta_B \bar m_\mu Y_\nu^B+\delta^\alpha_A\delta^\beta_u Y_\mu^A n_\nu+\delta^\alpha_A\delta^\beta_B(Y^A_\mu Y^B_\nu-\gamma^{AB}n_\mu m_\nu),\\
     A_{\mu\rho}^{\ \ \alpha}C_{\nu}^{\ \rho\beta}=&\ \delta^\alpha_r\delta^\beta_A \bar m_\mu Y^A_\nu-\delta^\alpha_A\delta^\beta_B(Y^A_\mu Y^B_\nu-\gamma^{AB}n_\mu m_\nu),\\
      A_{\mu\rho}^{\ \ \alpha}D_{\nu}^{\ \rho\beta A}=&\ { -n_\mu Y_\nu^A\delta^\alpha_r \delta^\beta_u+\bar{m}_\mu Y_\nu^A \delta^\alpha_r \delta^\beta_r -n_\mu n_\nu \gamma^{AB}\delta^\alpha_B\delta^\beta_u+(Y_\mu^BY_\nu^A+\gamma^{AB}n_\mu m_\nu)\delta^\alpha_{ B}\delta^\beta_r}\nn\\&+{ (\gamma^{BC}Y_\nu^An_\mu-\gamma^{AB}Y_\nu^C n_\mu)\delta^\alpha_B\delta^\beta_C},\\
     B_{\mu\rho}^{\ \ \alpha}B_{\nu}^{\ \rho\beta}=&\
     { -\frac{1}{2}(n_\mu \bar{n}_\nu+n_\nu \bar{n}_\mu )\delta^\alpha_u\delta_u^\beta+\bar{m}_\mu Y_\nu^B\delta_B^\beta\delta_u^\alpha+\bar{m}_\nu Y_\mu^B\delta_B^\alpha\delta_u^\beta+(Y_\mu^AY_\nu^B+\gamma^{AB}m_\mu m_\nu)\delta_A^\alpha\delta_B^\beta},\\
     B_{\mu\rho}^{\ \ \alpha}C_{\nu}^{\ \rho\beta}=&\ {-\bar m_\mu Y_\nu^A\delta^\alpha_u \delta^\beta_A-(Y_\mu^AY_\nu^B+\gamma^{AB}m_\mu m_\nu)\delta^\alpha_A\delta^\beta_B}\\
      B_{\mu\rho}^{\ \ \alpha}D_{\nu}^{\ \rho\beta A}=&\ { -n_\mu Y_\nu^A\delta^\alpha_u \delta^\beta_u+\bar{m}_\mu Y_\nu^A\delta^\alpha_u\delta^\beta_r-(Y_\nu^AY_\mu^B+\gamma^{AB}n_\nu m_\mu)\delta^\alpha_B\delta^\beta_u+(Y_\mu^BY_\nu^A+\gamma^{AB}m_\mu m_\nu)\delta^\alpha_B\delta^\beta_r}\nn\\&+{ (\gamma^{BC}Y_\nu^Am_\mu-\gamma^{AB}Y_\nu^C m_\mu)\delta^\alpha_B\delta^\beta_C}\\
     C_{\mu\rho}^{\ \ \alpha}C_{\nu}^{\ \rho\beta}=&\
     {
     (Y_\mu ^AY_\nu^B+\gamma^{AB}m_\mu m_\nu )\delta_A^\alpha\delta_B^\beta},\\
      C_{\mu\rho}^{\ \ \alpha}D_{\nu}^{\ \rho\beta A}=&\ 
      {
      (Y^B_\mu Y^A_\nu +\gamma^{AB}n_\nu m_\mu) \delta^\alpha_B\delta^\beta_u-(Y^B_\mu Y^A_\nu+\gamma^{AB}m_\mu m_\nu)\delta_B^\alpha\delta_r^\beta+(\gamma^{AB}Y^C_\nu m_\mu -\gamma^{BC}Y^A_\nu m_\mu )\delta_B^\alpha\delta_C^\beta},\\
      D_{\mu\rho}^{\ \ \alpha A}D_{\nu}^{\ \rho\beta B}=&\ { \gamma^{AB}n_\mu n_\nu \delta_u^\alpha\delta_u^\beta-(Y^A_\mu Y^B_\nu +\gamma^{AB}n_\mu m_\nu)\delta_u^\alpha\delta_r^\beta+(\gamma^{AB}Y_\nu^Dn_\mu -\gamma^{AD}Y^B_\nu n_\mu )\delta_u^\alpha\delta_D^\beta}\nn\\
      &{
      -(Y^A_\mu Y^B_\nu +\gamma^{AB}n_\nu m_\mu )\delta^\alpha_r\delta^\beta_u+(Y^A_\mu Y^B_\nu +\gamma^{AB}m_\mu m_\nu )\delta^\alpha_r\delta^\beta_r+(\gamma^{DA}Y^B_\nu m_\mu -\gamma^{AB}Y^D_\nu m_\mu )\delta^\alpha_r\delta^\beta_D}\nn\\
      &{
      +(\gamma^{AB}Y^C_\mu n_\nu       -\gamma^{BC}Y^A_\mu n_\nu )\delta_C^\alpha\delta_u^\beta+(-\gamma^{AB}Y^C_\mu m_\nu +\gamma^{BC}Y^A_\mu m_\nu)\delta^\alpha_C\delta^\beta_r}\nn\\
    & + { (\gamma^{CD}Y^A_\mu Y^B_\nu -\gamma^{BC}Y^A_\mu Y^D_\nu -\gamma^{AD}Y^C_\mu Y^B_\nu +\gamma^{AB}Y^C_\mu Y^D_\nu )\delta_C^\alpha\delta_D^\beta},
 \end{align}\es from which we obtain the form of $T_{\mu\nu}^{(k)}$
 \bea 
  T_{\mu\nu}^{(k)}&=&\sum_{m=0}^k\Bigg[
 -\frac{1}{2}(n_\mu \bar{n}_\nu +n_\nu \bar{n}_\mu)\dot{A}_r^{(m)}\dot{A}_r^{(k-m)}+(n_\mu Y^B_\nu+n_\nu Y^B_\mu) \dot{A}^{(m)}_r\dot{A}^{(k-m)}_B
+\gamma^{AB}n_\mu n_\nu\dot{A}^{(m)}_A\dot{A}^{(k-m)}_B\nn\\&&+(\Delta+k-m-1)[-(n_\mu \bar{n}_\nu +n_\nu \bar{n}_\mu)\dot{A}^{(m)}_rA^{(k-m-1)}_u   +(\bar{m}_\mu Y^B_\nu+\bar m_\nu Y_\mu^B)  \dot{A}^{(m)}_rA^{(k-m-1)}_B\nn\\&&+(Y_\mu ^A n_\nu+Y_\nu^A n_\mu) \dot{A}^{(m)}_AA^{(k-m-1)}_u  +(Y_\mu ^A Y_\nu ^B+Y_\nu^AY_\mu^B-\gamma^{AB}n_\mu m_\nu-\gamma^{AB}n_\nu m_\mu) \dot{A}^{(m)}_AA^{(k-m-1)}_B ]\nn\\
 &&+(\bar{m}_\mu Y^A_\nu+\bar m_\nu Y_\mu^A)\dot{A}^{(m)}_rA^{(k-m-1)}_A+(\gamma^{AB}(n_\mu m_\nu+n_\nu m_\mu) -Y_\mu ^A Y_\nu ^B-Y_\nu^AY_\mu^B)\dot{A}^{(m)}_AA^{(k-m-1)}_B\nn\\&&-(n_\mu Y_\nu^A+n_\nu Y^A_\mu)\dot{A}_r^{(m)}\nabla_A A_u^{(k-m-1)}+(\bar{m}_\mu Y_\nu ^A+\bar m_\nu Y_\mu^A)\dot{A}^{(m)}_r\nabla_AA^{(k-m-1)}_r\nn\\
 &&-2n_\mu n_\nu \dot{A}^{(m)}_A\nabla^AA^{(k-m-1)}_u+(Y_\mu ^A Y_\nu ^B+Y_\nu^AY_\mu^B+\gamma^{AB}n_\mu m_\nu +\gamma^{AB}n_\nu m_\mu)\dot{A}^{(m)}_B\nabla_AA^{(k-m-1)}_r\nn\\
 &&+(\gamma^{BC}Y_\nu ^An_\mu+\gamma^{BC}Y_\mu ^An_\nu -\gamma^{AB}Y^C_\nu n_\mu-\gamma^{AB}Y^C_\mu n_\nu )\dot{A}^{(m)}_B\nabla_AA^{(k-m-1)}_C\nn\\&&
 +(\Delta+m-1)(\Delta+k-m-1)[-\frac{1}{2}(n_\mu \bar{n}_\nu +n_\nu \bar{n}_\mu)A^{(m-1)}_uA^{(k-m-1)}_u+\bar{m}_\mu Y^B_\nu A^{(m-1)}_uA^{(k-m-1)}_B\nn\\
 &&+\bar{m}_\nu Y_\mu ^BA^{(m-1)}_BA^{(k-m-1)}_u+(Y^A_\mu Y^B_\nu +\gamma^{AB}m_\mu m_\nu )A^{(m-1)}_AA^{(k-m-1)}_B
 ]+(\Delta+m-1)[\nn\\
 &&-(\bar{m}_\mu Y^A_\nu+\bar m_\nu Y^A_\mu) A^{(m-1)}_uA^{(k-m-1)}_A-(Y_\mu ^A Y_\nu ^B+Y_\nu ^A Y_\mu ^B+ 2\gamma^{AB}m_\mu m_\nu )A^{(m-1)}_AA^{(k-m-1)}_B\nn\\&&-(n_\mu Y_\nu^A+n_\nu Y_\mu^A)A_u^{(m-1)}\nabla_AA_u^{(k-m-1)}+(\bar{m}_\mu Y_\nu ^A+\bar m_\nu Y_\mu^A)A^{(m-1)}_u\nabla_AA^{(k-m-1)}_r\nn\\
 &&-(Y_\mu ^BY_\nu ^A+Y_\nu ^BY_\mu ^A+\gamma^{AB}n_\nu m_\mu+\gamma^{AB}n_\mu m_\nu )A^{(m-1)}_B\nabla_AA^{(k-m-1)}_u\nn\\&&+(Y_\mu ^BY_\nu ^A+Y_\nu ^BY_\mu ^A+2\gamma^{AB}m_\mu m_\nu )A^{(m-1)}_B\nabla_AA^{(k-m-1)}_r\nn\\
 &&+(\gamma^{BC}Y^A_\nu m_\mu+\gamma^{BC}Y^A_\mu m_\nu -\gamma^{AB}Y^C_\nu m_\mu-\gamma^{AB}Y^C_\mu m_\nu )A^{(m-1)}_B\nabla_AA^{(k-m-1)}_C
]\nn\\&&
+(Y^A_\mu Y^B_\nu +\gamma^{AB}m_\mu m_\nu )A^{(m-1)}_AA^{(k-m-1)}_B\nn\\
 &&+(Y^B_\mu Y^A_\nu+Y^B_\nu Y^A_\mu  +\gamma^{AB}n_\nu m_\mu+\gamma^{AB}n_\mu m_\nu )A^{(m-1)}_B\nabla_AA^{(k-m-1)}_u\nn\\&&-(Y^B_\mu Y^A_\nu +Y^B_\nu Y^A_\mu+2\gamma^{AB}m_\mu m_\nu )A^{(m-1)}_B\nabla_AA^{(k-m-1)}_r\nn\\
 &&+(\gamma^{AB}Y^C_\nu m_\mu +\gamma^{AB}Y^C_\mu m_\nu-\gamma^{BC}Y_\mu ^Am_\nu -\gamma^{BC}Y_\nu ^Am_\mu )A^{(m-1)}_B\nabla_AA^{(k-m-1)}_C\nn\\&&
 +n_\mu n_\nu \nabla_AA_u^{(m-1)}\nabla^AA_u^{(k-m-1)}-(Y^A_\mu Y^B_\nu+Y^A_\nu Y^B_\mu +\gamma^{AB}n_\mu m_\nu +\gamma^{AB}n_\nu m_\mu)\nabla_AA_u^{(m-1)}\nabla_BA_r^{(k-m-1)}\nn\\&&+(\gamma^{AB}Y^D_\nu n_\mu+\gamma^{AB}Y^D_\mu n_\nu
     -\gamma^{AD}Y^B_\nu n_\mu-\gamma^{AD}Y^B_\mu n_\nu )\nabla_AA_u^{(m-1)}\nabla_BA_D^{(k-m-1)}\nn\\&&
     +(Y^A_\mu Y^B_\nu+\gamma^{AB}m_\mu m_\nu )\nabla_AA_r^{(m-1)}\nabla_BA_r^{(k-m-1)}\nn\\&&+(\gamma^{DA}Y^B_\nu m_\mu+\gamma^{DA}Y^B_\mu m_\nu -\gamma^{AB}Y^D_\nu m_\mu-\gamma^{AB}Y^D_\mu m_\nu )\nabla_AA_r^{(m-1)}\nabla_BA_D^{(k-m-1)}\nn\\&&
 +(\gamma^{CD}Y^A_\mu Y^B_\nu -\gamma^{BC}Y^A_\mu Y^D_\nu -\gamma^{AD}Y^C_\mu Y^B_\nu +\gamma^{AB}Y^C_\mu Y^D_\nu )\nabla_AA_C^{(m-1)}\nabla_BA_D^{(k-m-1)} \Bigg]\nn\\&&-\frac{1}{4}\eta_{\mu \nu}\text{trace}.\label{stressk}
 \eea

\section{Commutators}\label{comappendix}
In this Appendix, we will present some technical aspects of the commutators. Firstly, we will prove the formula \eqref{o12}. The starting point is \eqref{olambda} where we may write 
\be 
\bm g(\bm Y,\bm h)=\bm Y(\bm h)-\frac{1}{2}[\bm h,d\bm Y].
\ee We have defined a 2-form field from $\bm Y$ and $\bm h$
\bea 
\bm Y(\bm h)=\frac{1}{2}Y^C\nabla_Ch_{AB} d\theta^A\wedge d\theta^B\quad\Rightarrow\quad \left(\bm Y(\bm h)\right)_{AB}=Y^C\nabla_Ch_{AB}.
\eea 
Then \eqref{olambda} is 
\bea 
\bm o^{(\lambda)}(\bm Y,\bm Z)&=&\bm o(\bm Y,\bm Z)-\lambda d[\bm Y,\bm Z]+\lambda \bm g(\bm Y,d\bm Z)-\lambda \bm g(\bm Z,d\bm Y)-\lambda^2[d\bm Y,d\bm Z]\nn\\&=&\bm o(\bm Y,\bm Z)-\lambda d[\bm Y,\bm Z]+\lambda \bm Y(d\bm Z)-\lambda \bm Z(d\bm Y)+(\lambda-\lambda^2)[d\bm Y,d\bm Z].
\eea Now we just need to prove the following identity 
\bea 
d[\bm Y,\bm Z]-\bm Y(d\bm Z)+\bm Z(d\bm Y)-2\bm o(\bm Y,\bm Z)-\frac{1}{2}[d\bm Y,d\bm Z]=2\bm R(\bullet,\bullet,\bm Y,\bm Z)\label{dydz}
\eea with $\bm R$  the Riemann curvature tensor. 
\paragraph{Proof.}
In components, 
\bea 
\text{LHS}&=&\nabla_A[\bm Y,\bm Z]_B-\nabla_B[\bm Y,\bm Z]_A-Y^C\nabla_C(d\bm Z)_{AB}+Z^C\nabla_C(d\bm Y)_{AB}-2o_{AB}(\bm Y,\bm Z)-\frac{1}{2}[d\bm Y,d\bm Z]_{AB}\nn\\&=&\nabla_A(Y^C\nabla_CZ_B-Z^C\nabla_CY_B)-\nabla_B(Y^C\nabla_CZ_A-Z^C\nabla_CY_A)\nn\\&&-Y^C\nabla_C(\nabla_AZ_B-\nabla_BZ_A)+Z^C\nabla_C(\nabla_AY_B-\nabla_BY_A)\nn\\&&-\frac{1}{2}\Theta_{AC}(\bm Y)\Theta^C_{\ B}(\bm Z)+\frac{1}{2}\Theta_{AC}(\bm Z)\Theta^C_{\ B}(\bm Y)-\frac{1}{2}(d\bm Y)_{AC}(d\bm Z)^{C}_{\ B}+\frac{1}{2}(d\bm Z)_{AC}(d\bm Y)^C_{\ B}\nn\\&=&Y^C[\nabla_A,\nabla_C]Z_B-Z^C[\nabla_A,\nabla_C]Y_B-Y^C[\nabla_B,\nabla_C]Z_A+Z^C[\nabla_B,\nabla_C]Y_A\nn\\&&+\nabla_AY^C \nabla_CZ_B-\nabla_AZ^C \nabla_CY_B-\nabla_BY^C \nabla_CZ_A+\nabla_BZ^C\nabla_CY_A\nn\\&&-\frac{1}{2}(\nabla_AY_C+\nabla_CY_A)(\nabla^CZ_B+\nabla_BZ^C)+\frac{1}{2}(\nabla_AZ_C+\nabla_CZ_A)(\nabla^CY_B+\nabla_BY^C)\nn\\&&-\frac{1}{2}(\nabla_AY_C-\nabla_CY_A)(\nabla^CZ_B-\nabla_BZ^C)+\frac{1}{2}(\nabla_AZ_C-\nabla_CZ_A)(\nabla^CY_B-\nabla_BY^C)\nn\\&=&Y^CR_{BDAC}Z^D-Z^CR_{BDAC}Y^D-Y^CR_{ADBC}Z^D+Z^CR_{ADBC}Y^D\nn\\&=&2Y^CZ^D(R_{ACBD}-R_{BCAD})\nn\\&=&2R_{ABCD}Y^CZ^D\nn\\&=&\text{RHS}.
\eea We have used the definition 
\bea 
[\nabla_C,\nabla_D]Y_A=R_{ABCD}Y^B,
\eea the first Bianchi identity
\bea 
R_{ABCD}+R_{ACDB}+R_{ADBC}=0,
\eea the skew symmetry 
\bea 
R_{ABCD}=-R_{BACD},\quad R_{ABCD}=-R_{ABDC}
\eea and the  interchangeable symmetry of the Riemann tensor 
\be 
R_{ABCD}=R_{CDAB}.
\ee 

Using the identity \eqref{dydz}, we find 
\bea 
\bm o^{(\lambda)}(\bm Y,\bm Z)=(1-2\lambda)\bm o(\bm Y,\bm Z)-\lambda(\frac{1}{2}-\lambda)[d\bm Y,d\bm Z]-2\lambda\bm R(\bullet,\bullet,\bm Y,\bm Z).
\eea This is dramatically simplified for $\lambda=\frac{1}{2}$ 
\bea 
\bm o^{(1/2)}(\bm Y,\bm Z)=-\bm R(\bullet,\bullet,\bm Y,\bm Z).
\eea 

Secondly, we will  check the Jacobi identities associated with the commutators. The Jacobi identity 
\bea 
[[\mathcal{M}^{(1/2)}_{\bm X},\mathcal{M}^{(1/2)}_{\bm Y}],\mathcal{M}^{(1/2)}_{\bm Z}]+[[\mathcal{M}^{(1/2)}_{\bm Y},\mathcal{M}^{(1/2)}_{\bm Z}],\mathcal{M}^{(1/2)}_{\bm X}]+[[\mathcal{M}^{(1/2)}_{\bm Z},\mathcal{M}^{(1/2)}_{\bm X}],\mathcal{M}^{(1/2)}_{\bm Y}]=0
\eea is satisfied due to the previous properties of the Riemann curvature tensor and the second Bianchi identity 
\be 
\nabla_ER_{CDAB}+\nabla_CR_{DEAB}+\nabla_{D}R_{ECAB}=0.
\ee Similarly, the Jacobi identity 
\bea 
[[\mathcal{M}^{(1/2)}_{\bm Y},\mathcal{M}^{(1/2)}_{\bm Z}],\mathcal{O}^{(1/2)}_{\bm h}]+[[\mathcal{M}^{(1/2)}_{\bm Z},\mathcal{O}^{(1/2)}_{\bm h}],\mathcal{M}^{(1/2)}_{\bm Y}]+[[\mathcal{O}^{(1/2)}_{\bm h},\mathcal{M}^{(1/2)}_{\bm Y}],\mathcal{M}^{(1/2)}_{\bm Z}]=0
\eea  can be checked with the identity 
\bea 
[\nabla_C,\nabla_D]h_{AB}=R_{AECD}h^E_{\ B}+R_{ABCD}h_A^{\ E}.
\eea 

\section{General Carrollian manifold}\label{hyper}
This section is a collection of the result of the vector theory on a general Carrollian manifold with topology $\mathcal{N}=\mathbb{R}\times N$. The scalar theory on such manifold has been studied in \cite{Li:2023xrr}. The rigorous discussion on Carrollian manifold can be found in \cite{Duval_2014a,Duval_2014b,Duval:2014uoa,Ciambelli:2018xat,Ciambelli:2018wre,Ciambelli:2019lap} and the relation between Carrollian holography and celestial holography has been discussed in \cite{Donnay:2022aba,Bagchi:2022emh,Donnay:2022wvx}. We divide the discussion into two parts. The first part is an intrinsic derivation of the supertranslation, superrotation and superduality transformation while the second part
will use the method of bulk reduction.

\subsection{Intrinsic derivation}
The metric of the Riemann manifold $N$ is
\be 
ds_N^2=\gamma_{AB}d\theta^Ad\theta^B,\quad A=1,2,\cdots, m,
\ee where the coordinates $\theta^A$ are not necessary the spherical coordinates of $ S^m$.
We will leave the metric $\gamma_{AB}$ free except that its determinant is non-zero such that there is an inverse metric matrix $\gamma^{AB}$. The Carrollian manifold $\mathcal{N}=\mathbb{R}\times N$ is described by $d-1$ coordinates $(u,\Omega)=(u,\theta^A)$ where $u$ can be interpreted as a time parameter. The fundamental vector field on $\mathcal{N}$ is $A_A(u,\Omega)$ and the symplectic form is 
\bea 
\bm\Omega(\delta A;\delta A;A)={ -}\int du d\Omega \delta\dot{A}^A\wedge\delta A_A.\label{symp}
\eea We will assume the  variations of the field $A_A$ are 
\bs\begin{align}\slashed\delta_f A_A=&\Delta_A(f;A;u,\Omega)=f(u,\Omega)\dot{A}_A(u,\Omega),\label{st1}\\
\hspace{-2em}\slashed\delta_{\bm Y} A_A=&\Delta_A(Y;A;u,\Omega)=Y^C\nabla_CA_A+\frac{1}{2}\nabla_CY^C A_A+\frac{1}{2}(\nabla_AY_C-\nabla_CY_A)A^C\label{sr1}
\end{align}\es under supertranslation and superrotation respectively. It follows immediately that the corresponding  Hamiltonians are still $\mathcal{T}_f$ and $\mathcal{M}_{\bm Y}$  and the commutators obey the same form as \eqref{commutators} from which we read out the helicity flux operator $\mathcal{O}_{\bm h}$ whose form is still \eqref{helicityfluxoperator}. It is clear that the derivation is independent of the explicit metric of the Riemann manifold $N$ and the result is universal for any dimensions.

\subsection{Carrollian manifold as a null hypersurface}
Now we will try to embed the Carrollian manifold $\mathcal{N}$ into a higher dimensional spacetime whose metric in null Gaussian coordinate system is \cite{Moncrief:1983xua,chrusciel2020geometry,2019CQGra..36p5002D}
\bea 
ds^2=K du^2-2du d\rho+H_{AB}(d\theta^A+\Lambda^A du)(d\theta^B+\Lambda^B du)\label{embmetric}
\eea where $K,\Lambda^A,H_{AB}$ depend on the coordinates $u,\rho,\theta^A$. The Carrollian manifold is a co-dimension one null hypersurface which is located at $\rho=0$ and therefore we may assume the asymptotic expansion near $\rho=0$ 
\bs\begin{align} 
K(u,\rho,\theta)&=-2\kappa(u,\theta)\rho-2\sum_{k=2}^\infty \kappa^{(k)}(u,\theta)\rho^k,\\
\Lambda^A(u,\rho,\theta)&=\lambda^A(u,\theta)\rho+\sum_{k=2}^\infty \lambda^{A(k)}(u,\theta)\rho^k,\\
H_{AB}(u,\rho,\theta)&=\gamma_{AB}+\sum_{k=1}^\infty \gamma_{AB}^{(k)}(u,\theta)\rho^k.
\end{align}\es 
 The components of the metric could be written out explicitly as
\bs\begin{align}
    g_{uu}&=K+H_{AB}\Lambda^A\Lambda^B=-2\kappa\rho+\mathcal{O}(\rho^2),\\
    g_{u\rho}&=-1,\quad g_{\rho\rho}=0,\quad g_{\rho A}=0,\\
    g_{uA}&=H_{AB}\Lambda^B=\lambda_{A}\rho+\mathcal{O}(\rho^2),\\
    g_{AB}&=H_{AB}=\gamma_{AB}+\mathcal{O}(\rho).
\end{align}\es  The  components of the inverse metric is \bs\begin{align}
    g^{uu}&=0,\quad g^{u\rho}=-1,\quad g^{uA}=0,\\
    g^{\rho\rho}&=-K=2\kappa\rho+\mathcal{O}(\rho^2),\\ g^{\rho A}&=H^{AB}\Lambda_B=\lambda^A \rho+\mathcal{O}(\rho^2),\\
    g^{AB}&=H^{AB}= \gamma^{AB}+\mathcal{O}(\rho).
\end{align}\es We will consider a vector theory whose action is still \eqref{actionvector}. However, the fall-off condition for the vector field becomes 
\be 
a_u=\sum_{k=0}^\infty A_u^{(k)}(u,\Omega)\rho^k,\quad a_A=\sum_{k=0}^\infty A_A^{(k)}(u,\Omega)\rho^k
\ee under the radial gauge 
\be 
a_\rho=0.
\ee The components of the electromagnetic field 
 are 
 \bs\begin{align}
     f_{u\rho}&=-\sum_{k=1}^\infty k A_u^{(k)}\rho^{k-1}=-A_u^{(1)}+\mathcal{O}(\rho),\\
     f_{uA}&=\sum_{k=0}^\infty (\dot{A}_A^{(k)}-\partial_A A_u^{(k)})\rho^k=(\dot{A}_A-\partial_AA_u)+\mathcal{O}(\rho),\\
     f_{\rho A}&=\sum_{k=1}^\infty k A_A^{(k)}\rho^{k-1}=A_A^{(1)}+\mathcal{O}(\rho),\\
     f_{AB}&=\sum_{k=0}^\infty (\partial_AA_B^{(k)}-\partial_BA_A^{(k)})\rho^k=(\partial_AA_B-\partial_BA_A)+\mathcal{O}(\rho).
 \end{align}\es The contravariant vector is 
 \bs\begin{align}
     a^u=&0,\\ a^\rho=&-a_u+H^{AB}\Lambda_Ba_A=-A_u-(A_u^{(1)}-\lambda^A A_A)\rho+\mathcal{O}(\rho^2),\\
     a^A=&H^{AB}a_B=A^A+\mathcal{O}(\rho). 
 \end{align}\es
 The contravariant electromagnetic field is
 \bs\begin{align}
     f^{u\rho}&=A_u^{(1)}+\mathcal{O}(\rho),\\
     f^{u A}&=-\gamma^{AB}A_B^{(1)}+\mathcal{O}(\rho),\\
     f^{\rho A}&=-(\dot{A}^A-\partial^AA_u)+\mathcal{O}(\rho),\\
     f^{AB}&=\nabla^AA^B-\nabla^BA^A+\mathcal{O}(\rho).
 \end{align}\es  The symplectic form is 
\begin{align}
  \bm\Omega(\delta A;\delta A;A)&=-\lim_{\rho\to 0}\int_{\ch_{\rho}} (d^{d-1}x)_\mu \delta f^{\mn}\wedge \delta a_\nu\nn\\
  &=\lim_{\rho\to 0}\int_{\ch_{\rho}} du d\Omega (\delta f^{\rho u}\wedge \delta a_u+\delta f^{\rho A}\wedge \delta a_A)\nn\\
  &=\int du d\Omega (-\delta A_u^{(1)}\wedge \delta A_u+\delta {A}^A\wedge\delta \dot A_A+\partial^A\delta A_u\wedge \delta A_A),
\end{align}
where we have used the volume form of a constant $\rho$ hypersurface $\ch_{\rho}$
\begin{align}
  (d^{d-1}x)_\mu=-\sqrt{2\kappa\rho}\,\mathfrak{m}_\mu du d\Omega=-\delta_\mu^\rho dud\Omega.
\end{align}
 
 We may impose a further condition 
 \be 
A_u=0 \label{Au0}
 \ee to simplify the symplectic form.
 The reason is  shown as follows. 
 \begin{enumerate}
     \item For the Rindler horizon, we can find this condition from the standard mode expansion of plane waves from taking the limit approaching the Rindler horizon in the standard mode expansion of the vector field.
     \item Under this condition, we can find 
 \be 
 \bm\Omega(\delta A;\delta A;A)=\int du d\Omega \ \delta A_A\wedge \delta\dot{A}^A
 \ee which is the same as \eqref{symp}.
 \end{enumerate}
 From the $\rho$ component of the equation of motion
 \be 
 \nabla_\mu f^{\mu\nu}=0,
 \ee we find 
 \be 
 \dot{A}_u^{(1)}+\nabla_A(\dot{A}^A-\nabla^A A_u)=0.\label{Au1}
 \ee The  equation \eqref{Au1} is solved by 
 \be 
 A_u^{(1)}=-\nabla_AA^A+\varphi(\Omega).\label{Au1solution}
 \ee

Now we will show that the fluxes associated with the Carrollian diffeomorphism are exactly the Hamiltonians at the boundary. The leading order of the stress tensor is 
\be 
T^{\mu{(0)}}_{\ 
\ \nu}=f^{\mu\lambda(0)}f_{\nu \lambda(0)}-\frac{1}{4}\delta^\mu_\nu f^{(0)}_{\lambda\zeta}f^{\lambda\zeta(0)}
\ee 
whose components can be obtained from
\bs\begin{align}
    f^{u\lambda(0)}f_{u\lambda(0)}&=-\left(A_u^{(1)}\right)^2-A^{A(1)}(\dot A_A-\partial_AA_u),\\
   f^{u\lambda(0)}f_{\rho\lambda(0)}&=-A^{A(1)}A_A^{(1)},\\
    f^{u\lambda(0)}f_{A\lambda(0)}&=A_u^{(1)}A_A^{(1)}{ -}A^{B(1)}(\partial_AA_B-\partial_BA_A),\\
    f^{\rho\lambda(0)}f_{u\lambda(0)}&=-(\dot A^A-\partial^AA_u)(\dot A_A-\partial_AA_u),\\
    f^{\rho\lambda(0)}f_{\rho\lambda(0)}&=-\left(A_u^{(1)}\right)^2-A^{A(1)}(\dot A_A-\partial_AA_u),\\
    f^{\rho\lambda(0)}f_{A\lambda(0)}&=A_u^{(1)}(\dot A_A-\partial_AA_u)-(\dot A^B-\partial^BA_u)(\partial_AA_B-\partial_BA_A),\\
    f^{A\lambda(0)}f_{u\lambda(0)}&=-A_u^{(1)}(\dot A^A-\partial^AA_u)+(\partial^AA^B-\partial^BA^A)(\dot A_B-\partial_BA_u),\\
    f^{A\lambda(0)}f_{\rho\lambda(0)}&= A_u^{(1)}\dot{A}^{A(1)} +(\partial^AA^B-\partial^BA^A)A_B^{(1)},\\
    f^{A\lambda(0)}f_{B\lambda(0)}&= -A^{A(1)}(\dot A_B-\partial_BA_u) -(\dot A^A-\partial^AA_u)A_B^{(1)}+(\nabla^AA^C-\nabla^CA^A)(\nabla_BA_C-\nabla_CA_B).
\end{align}
\es 
Using the condition \eqref{Au0} and the solution \eqref{Au1solution}, we find the relevant components of the stress tensor 
\bs\begin{align}
    T^\rho_{\ u}&=-\dot A^A\dot A_A,\\
    T^\rho_{\ A}&=-\dot A_A \nabla_BA^B-\dot A^B(\nabla_AA_B-\nabla_BA_A).
\end{align}
\es 
The Carrollian diffeomorphism is generated by the vector 
\be 
\bm\xi=f(u,\Omega)\partial_u+Y^A(\Omega)\partial_A.
\ee Therefore, the corresponding fluxes from bulk to boundary  are
\bs\begin{align} 
\mathcal{Q}_f&=\int du d\Omega T^\rho_{\ u}\xi^u=-\int du d\Omega f(u,\Omega)\dot A^A\dot A_A,\\
\mathcal{Q}_{\bm Y}&=\int du d\Omega T^\rho_{\ A}\xi^A=\int du d\Omega Y^A(\Omega)(-\dot A_A\nabla_BA^B-\dot A^B(\nabla_AA_B-\nabla_BA_A)).
\end{align}\es 
 These are consistent with the form of the corresponding Hamiltonians at the boundary.
 
 Now at the null boundary, we may find the following vector field 
 \be 
 a=A_u du+A_A d\theta^A=A_Ad\theta^A.
 \ee At the second step, we have used the condition \eqref{Au0}. Now we can evaluate the Chern-Simons term in three dimensional Carrollian manifold 
 \bea 
 I[a]=\int a\wedge da=\int A_A d\theta^A \wedge \dot{A}_B du\wedge d\theta^B=-\int du d\Omega \epsilon^{AB}A_A\dot{A}_B.
 \eea This is exactly the minimal helicity flux operator. Now we can also consider the $d-1$ dimensional Carrollian manifold ($d>4$)
 \bea 
 I[a]&=&\int a\wedge da\wedge {\tt g}\nn\\&=&\int A_A d\theta^A \wedge \dot{A}_B du \wedge d\theta^B\wedge \frac{1}{(d-4)!}{\tt g}_{C_1\cdots C_{d-4}}d\theta^{C_1}\wedge\cdots\wedge d\theta^{C_{d-4}}\nn\\&&+\int A_A d\theta^A \wedge \partial_B A_C d\theta^B\wedge d\theta^C\wedge{\tt g}_{u\cdots} du\wedge\cdots\nn\\&=&-\int du d\Omega A_A\dot{A}_B h^{AB}.\label{CSd}
 \eea Note that in the last step, we have assumed the components ${\tt g}_{u\cdots}=0$. In \cite{Baulieu:1997nj}, the Chern-Simons  term $I[a]$ is an observable where $\tt g$ is a closed $d-4$ form coming from $S^{d-2}$, which leads exactly to the same condition.
 
 \section{Large gauge transformation}\label{gaugechoice}
 
\paragraph{Boundary aspect.}
In this Appendix, we will discuss the  residual gauge transformation 
\be A_A\to A_A+\partial_A\epsilon.\label{residuegauge}
\ee 
Utilizing Hamilton's equation, we can  derive the associated charge
\bea 
\delta H_\epsilon=\int du d\Omega \delta \dot{A}^A\partial_A\epsilon=\delta \int d\Omega \left(A^A(u,\Omega)\Big|_{-\infty}^\infty\right)\nabla_A\epsilon(\Omega).
\eea 
In the context, we have imposed a strong condition on the form 
\be 
q^A(\Omega)=A^A(u=\infty,\Omega)-A^A(u=-\infty,\Omega)
\ee such that $q^A=0$ and then $\mathcal{M}_{\bm Y}$ and $\mathcal{O}_{\bm h}$ are gauge invariant. However, we can also consider the possibility that $q^A\not=0$ such that the residual gauge transformation \eqref{residuegauge} becomes a large gauge transformation. 
This indicates a nontrivial integrable charge 
\be 
H_\epsilon=\int d\Omega\  q^A(\Omega)  \nabla_A\epsilon(\Omega)\label{integrablecharge}
\ee which is exactly the soft charge of \cite{Strominger:2017zoo} once the magnetic field $F_{AB}$ vanishes at $\mathcal{I}^+_{\pm}$
\be 
F_{AB}\Big|_{\mathcal{I}^+_{\pm}}=0.
\ee To see this point, we solve the above equation and define
\be 
A_A(u,\Omega)\Big|_{-\infty}^\infty=\nabla_A K(\Omega)\quad\Rightarrow\quad H_{\epsilon}=-\int d\Omega \epsilon \nabla^2 K(\Omega).\label{defK}
\ee The mode $K(\Omega)$ is soft and its commutator with the radiative mode $A_A(u,\Omega)$ is subtle since there would be a discrepancy of factor 2 compared to taking the limit of \eqref{comAA}.  This mismatch may be solved by taking into account the constraint  $F_{AB}=0$ at $\mathcal{I}^{+}_{\pm}$ and modifying the Poisson brackets \cite{He:2014laa,He:2014cra}.  Finally,  $H_{\epsilon}$ generates the large gauge transformation 
\be 
\ [H_{\epsilon},A_A(u,\Omega)]=-i\partial_A\epsilon(\Omega).
\ee  
Large gauge transformations are physical and then the variation $\delta_{\epsilon}\mathcal{M}_{\bm Y}$ and $\delta_{\epsilon}\mathcal{O}_{\bm h}$ can be non-zero. 
 Interestingly, we can compute the following commutator 
 \bea 
 \ [\mathcal{O}_{\bm h},H_\epsilon]=i\int d\Omega q^A(\Omega) h_{BA}(\Omega)\partial^B\epsilon(\Omega)=-i\int d\Omega K(\Omega)\nabla^A h_{BA}(\Omega)\nabla^B\epsilon=iH_{\tilde{\epsilon}_1} \label{magcharge}
 \eea and the right-hand side is still the operator $H_{\tilde{\epsilon}_1}$ where $\tilde{\epsilon}_1$ obeys the equation \be \nabla^B h_{BA} \nabla^A\epsilon=\nabla^2\tilde{\epsilon}_1.\ee 
 Similarly, we can compute the commutator
 \bea 
 \ [\mathcal{M}_{\bm Y}, H_{\epsilon}]&=&i\int d\Omega q^A [Y^C\nabla_C\nabla_A\epsilon+\frac{1}{2}\nabla_CY^C\nabla_A\epsilon+\frac{1}{2}(\nabla_AY_B-\nabla_BY_A)\nabla^B\epsilon]\nn\\&=&-i \int d\Omega K(\Omega)\nabla^A[Y^C\nabla_C\nabla_A\epsilon+\frac{1}{2}\nabla_CY^C\nabla_A\epsilon+\frac{1}{2}(\nabla_AY_B-\nabla_BY_A)\nabla^B\epsilon]\nn\\&=&i H_{\tilde{\epsilon}_2},
 \eea where $\tilde{\epsilon}_2$ is determined by 
 \bea 
 \nabla^2\tilde{\epsilon}_2=\nabla^A[Y^C\nabla_C\nabla_A\epsilon+\frac{1}{2}\nabla_CY^C\nabla_A\epsilon+\frac{1}{2}(\nabla_AY_B-\nabla_BY_A)\nabla^B\epsilon].
 \eea 
Finally, the commutator between $H_{\epsilon_1}$ and $H_{\epsilon_2}$ vanishes.
\paragraph{Bulk aspect.}
 One can also discuss the problem from the bulk side. The surface charge associated with the transformation \eqref{fallepsilon} is 
\bea 
\mathcal Q_{\epsilon}&=&-\lim{}\!_+\int \epsilon_{\text{bulk}}*f\nn\\&=&-\lim{}\!_+\int \epsilon_0 * f-\lim{}\!_+\int \left( \frac{\epsilon(\Omega)}{r^{\Delta-1}}+\cdots\right)\epsilon^{ur}_{\ \ A_1\cdots A_m}f_{ur}d\theta^{A_1}\wedge d\theta^{A_2}\cdots\wedge d\theta^{A_m}\label{surfacecharge}
\eea 
Note that the first term with $\epsilon_0$ is divergent superficially. However, the divergence can be removed by using the equations  of motion and Stokes' theorem \cite{Campoleoni:2017qot}. This term contributes to the electric charge and we will denote it as 
$\mathcal Q_{\epsilon}^{(1)}
$. The second term is more interesting, we rewrite it as 
\be 
\mathcal Q_{\epsilon}^{(2)}=\lim{}\!_+ r^{2\Delta}\int d\Omega \frac{\epsilon(\Omega)}{r^{\Delta-1}}f_{ur}.
\ee The $r^{2\Delta}$ is from the integral measure on the celestial sphere. The field $f_{ur}$ can be expanded near $\mathcal I^+$ as
\be 
f_{ur}=\frac{\dot A_r}{r^\Delta}+\frac{\dot A_r^{(1)}+\Delta A_u}{r^{\Delta+1}}+\cdots=\frac{\dot A_r^{(1)}+\Delta A_u}{r^{\Delta+1}}+\cdots.
\ee In the second step, we have used the equation of motion \eqref{dotAr0}. Therefore, the surface charge $\mathcal Q_{\epsilon}^{(2)}$ is also finite
\bea 
\mathcal Q_\epsilon^{(2)}(u)=\int d\Omega \epsilon(\Omega)\left(\nabla_A A^A+\widetilde{\varphi}(\Omega)\right)
\eea where we have used the equations of motion \eqref{Aueom} and \eqref{dotAr0}. Compared with \eqref{integrablecharge}, we find that the  difference of the surface charge between $\mathcal{I}^+_{+}$ and $\mathcal{I}_-^+$ is exactly the integrable flux $-H_{\epsilon}$
\bea 
\mathcal Q_{\epsilon}(u=+\infty)-\mathcal Q_{\epsilon}(u=-\infty)=-H_\epsilon.
\eea The time-independent part in $\mathcal Q_{\epsilon}(u)$ has been canceled exactly. This is consistent with the fact that the flux will contribute to the increment of the surface charge. 

\section{Helicity representation of Poincar\'e group in higher dimensions}\label{littlegroup}
The unitary representation of Poincar\'e group  in four dimensions has been established since the work of \cite{Wigner:1939cj,1948ZPhy..124..665W,Bargmann:1948ck} and it has been nicely reviewed in the classic book 
\cite{weinberg_1995}. We will collect the basic elements which are related to this work. Recent developments on the representation of higher dimensional Poincar\'e group include \cite{Bekaert:2006py,Weinberg:2020nsn,Kuzenko:2020ayk,Buchbinder:2020ocz}. The Poincar\'e group ${ISO}(1,d-1)$ are the semi-product of the spacetime translations and the Lorentz transformations. The Poincar\'e algebra $iso(1,d-1)$ is generated by the momentum $P_\mu$ and angular momentum $J_{\mu\nu}$ 
\bs\begin{align}
    [P_\mu,P_\nu]&=0,\\
    [J_{\mu\nu},P_\rho]&=i\eta_{\mu\rho}P_\nu-i\eta_{\nu\rho}P_\mu,\\
    [J_{\mu\nu},J_{\rho\sigma}]&=i(\eta_{\mu\rho}J_{\nu\sigma}-\eta_{\mu\sigma}J_{\nu\nu}-\eta_{\nu\rho}J_{\mu\sigma}+\eta_{\nu\sigma}J_{\mu\rho}).
\end{align}\es 
In four dimensions, there are two independent Casimir operators, $P^2=P_\mu P^\mu$ defines the mass $m$ while $W^2=W_\mu W^\mu$ defines the spin $s$ of the representation\footnote{This is only useful for massive representations. We will discuss the  massless representations later.}
\bea 
-P^2=m^2,\quad W^2=m^2 s(s+1).\label{massive}
\eea We have introduced the famous Pauli-Lubanski pseudo-vector  
\be 
W^\mu=\frac{1}{2}\epsilon^{\mu\nu\rho\sigma}J_{\nu\rho}P_\sigma.
\ee 
In higher dimensions, there is a similar generalized Pauli-Lubanski tensor 
\bea 
W^{\mu_1\cdots\mu_{d-3}}=\frac{1}{2}\epsilon^{\mu_1\cdots\mu_{d-3}\nu\rho\sigma}J_{\nu\rho}P_\sigma
\eea and the corresponding irreducible representations depend on the dimension. Nevertheless, one can focus on 
the orbits of the momentum and the irreducible representations can be classified into the massive, massless, tachyonic and zero-momentum representations whose little group can be found in \cite{Bekaert:2006py}.
\begin{table}
\begin{center}
\renewcommand\arraystretch{1.5}
 \begin{tabular}{|c||c|c|c|}\hline
\text{Momentum}&\text{Orbit}&\text{Little group}&\text{Unitary irreducible representation}\\\hline
$p=0$&\text{Origin}&$SO(1,d-1)$&\text{Zero momentum}\\\hline
$p^2=-m^2$&\text{Mass-shell}&$SO(d-1)$&\text{Massive}\\\hline
$p^2=0$&\text{Light cone}&$ISO(d-2)$&\text{Massless}\\ \hline
$p^2=m^2$& \text{Hyperboloid}& $SO(1,d-2)$& \text{Tachyonic}\\\hline
\end{tabular}
\caption{\centering{Orbits of the momentum}}\label{corres}
\end{center}
\end{table} 

The massless representation can be classified further according to the representations of the little group $ISO(d-2)$. This is a Euclidean group which is generated by $d-2$ ``momenta'' $\pi_a$ and $\frac{(d-2)(d-3)}{2}$ ``angular momenta''. and there are two possible orbits for the corresponding ``momentum''. The infinite spin representation has a non-vanishing ``momentum'' whose stability subgroup is $SO(d-3)$ while the helicity representation has a vanishing ``momentum'' whose stability subgroup is $SO(d-2)$. This is called the short little group and isomorphic to the group generated by the helicity flux operators with $\bm h$ obeys \eqref{hab}.

For the helicity representation, one can impose one more constraint
\bea 
P^{[\mu}W^{\nu_1\cdots\mu_{d-3}]}=0\label{pW0}
\eea besides the massless condition
\bea 
P^2=0.\label{P20}
\eea In four dimensions, the solution of \eqref{pW0} and \eqref{P20} is 
\bea 
W^\mu=\lambda P^\mu
\eea where $\lambda$ is the helicity. To find this result, one can choose an initial frame and set 
\bea 
P^\mu=E(1,0,0,1).
\eea Then the Pauli-Lubanski pseudo-vector is 
\bea 
W^0=-J_{12}E,\quad W^1=(J_{02}-J_{23})E,\quad W^2=(J_{13}-J_{01})E,\quad W^3=-J_{12}E.
\eea To satisfy the condition \eqref{pW0}, we find 
\bea 
W^1=W^2=0. 
\eea Therefore, the Pauli-Lubanski pseudo-vector is indeed proportional to the 4-momentum and the helicity $\lambda$ 
\bea 
\lambda=-J_{12}
\eea whose value takes 
\bea \lambda=0,\pm\frac{1}{2},\pm 1,\cdots.
\eea 
Utilizing the three-dimensional Levi-Civita tensor $\widetilde{\epsilon}^{ijk}$, we may define a pseudo-vector 
\bea 
S^i=\frac{1}{2}\widetilde{\epsilon}^{ijk}J_{jk}
\eea which is called spin in quantum mechanics. Then the helicity may be defined as the projection of the spin $\bm S$ into the direction of the 3-momentum 
\bea 
\lambda=-\frac{\bm S\cdot\bm P}{|\bm P|}=-\frac{S^i P^i}{|\bm P|}.
\eea 

In higher dimensions, we still choose an initial frame 
\bea 
P^\mu=E(1,0,\cdots,0,1).
\eea For later convenience, we denote the direction of the momentum $\bm P$ as $\hat{a}$ and the transverse directions as $a$, then 
\bea 
P^\mu=E\delta^\mu_0+E \delta^\mu_{\hat{a}}.
\eea The Pauli-Lubanski tensor becomes 
\bs\begin{align}
W_{0a_1\cdots a_{d-4}}&=\frac{1}{2}\epsilon_{a_1\cdots a_{d-4}bc}J^{bc}E,\\
W_{\hat{a}a_1\cdots a_{d-4}}&=-\frac{1}{2}\epsilon_{a_1\cdots a_{d-4}{bc}}J^{bc}E,\\
W_{a_1\cdots a_{d-5}0\hat{a}}&=0,\\
W_{a_1\cdots a_{d-3}}&=(-1)^d\epsilon_{a_1\cdots a_{d-3}b}(J^{\hat{a}b}-J^{0b})E.
\end{align}\es Note that $\epsilon_{a_1\cdots a_{d-2}}\equiv\epsilon_{0a_1\cdots a_{d-2}\hat a} $ is the Levi-Civita tensor of the Euclidean space $\mathbb{R}^{d-2}$ spanned by the $d-2$ transverse directions. The equation \eqref{pW0} is satisfied by the conditions
\bea 
J^{\hat{a}a}=J^{0a}\quad\Rightarrow\quad W_{a_1\cdots a_{d-3}}=0.
\eea Therefore, the Pauli-Lubanski tensor is also proportional to the momentum 
\bea
W^{0a_1\cdots a_{d-4}}=W^{\hat{a}a_1\cdots a_{d-4}}=\lambda^{a_1\cdots a_{d-4}}E,
\eea where we have defined a helicity tensor 
\bea 
\lambda_{a_1\cdots a_{d-4}}=-\frac{1}{2}\epsilon_{a_1\cdots a_{d-4}bc}J^{bc}.
\eea In terms of the $d-1$ dimensional Levi-Civita tensor $\widetilde{\epsilon}^{i_1\cdots i_{d-1}}\equiv \epsilon_{0}^{\ i_1\cdots i_{d-1}}$, we may define the following spin tensor
\bea 
S^{i_1\cdots i_{d-3}}=\frac{1}{2}\widetilde{\epsilon}^{i_1\cdots i_{d-3}jk}J_{jk}.
\eea Then the helicity tensor is still the projection of the spin tensor into the direction of the spatial momentum $\bm P$
\bea 
\lambda^{i_1\cdots i_{d-4}}=-\frac{S^{i_1\cdots i_{d-4}l}P_l}{|\bm P|}.
\eea This is a totally anti-symmetric tensor which is orthogonal to $\bm P$. Therefore, we may regard it as a totally anti-symmetric tensor in $\mathbb{R}^{d-2}$ whose 
 Hodge dual 
is a 2-form. The result is consistent with \cite{2024IJTP...63..149D} where the bundle structure of this massless helicity representation has been discussed and the helicity can be labeled by an anti-symmetric tensor which generates $SO(d-2)$. 
\bibliography{refs}
\end{document}